\RequirePackage{fix-cm}
\documentclass[smallextended]{svjour3}       % onecolumn (second format)
\smartqed  % flush right qed marks, e.g. at end of proof
\usepackage{graphicx}
\usepackage[dvipsnames]{xcolor}
\usepackage{tcolorbox}
\usepackage{float}
\usepackage{paralist}
\usepackage{rotating}
\usepackage{booktabs}
\usepackage{multirow}
\usepackage{listings}
\usepackage[a4paper]{geometry}
\usepackage{lscape}
\usepackage{ifthen}
\usepackage{amssymb}
\usepackage{enumitem}
\usepackage[caption=false]{subfig}
\usepackage{xspace}
\usepackage{fontawesome}
\usepackage{url}
\usepackage{longtable}
\usepackage[title]{appendix}

\definecolor{codegreen}{rgb}{0,0.6,0}
\definecolor{codegray}{rgb}{0.5,0.5,0.5}
\definecolor{codepurple}{rgb}{0.58,0,0.82}
\definecolor{backcolour}{rgb}{0.95,0.95,0.95}

\lstdefinestyle{mystyle}{
    backgroundcolor=\color{backcolour},   
    commentstyle=\color{codegreen},
    keywordstyle=\color{magenta},
    numberstyle=\tiny\color{codegray},
    stringstyle=\color{codepurple},
    basicstyle=\ttfamily\footnotesize,
    breakatwhitespace=false,         
    breaklines=true,                 
    captionpos=b,                    
    keepspaces=true,                 
    numbers=left,                    
    numbersep=5pt,                  
    showspaces=false,                
    showstringspaces=false,
    showtabs=false,                  
    tabsize=2,
}

\lstdefinestyle{mystyle2}{
    commentstyle=\color{OliveGreen},
    basicstyle=\itshape\footnotesize,
    breakatwhitespace=false,         
    breaklines=true,                 
    captionpos=b,                    
    keepspaces=true,                 
    numbers=none,
    showspaces=false,                
    showstringspaces=false,
    showtabs=false,                  
    tabsize=2,
    columns=fullflexible,
    frame=single,
    framerule=0.2pt
}

\begin{document}

\newboolean{showcomments}
\setboolean{showcomments}{true}
\ifthenelse{\boolean{showcomments}}
  {\newcommand{\nb}[2]{
    \fbox{\bfseries\sffamily\scriptsize#1}
    {\sf\small$\blacktriangleright$\textit{#2}$\blacktriangleleft$}
   }
  }
  {\newcommand{\nb}[2]{}
  }

\newcommand\TODO[1]{\textcolor{red}{\nb{TODO}{#1}}}

\newcommand\CSABA[1]{\textcolor{red}{\nb{CSABA}{#1}}}

\newcommand\ANTHONY[1]{\textcolor{red}{\nb{ANTHONY}{#1}}}

\newcommand\Foutse[1]{\textcolor{red}{\nb{Foutse: }{#1}}}

\newcommand\Biruk[1]{\textcolor{green}{\nb{Biruk: }{#1}}}

\newcommand\codecomment[1]{%
\begin{lstlisting}[language=Java]^^J%
// #1^^J
\end{lstlisting}
}

\newcommand{\deemph}[1]{{\color{black!40}#1}}

\newcommand{\ie}{\emph{i.e.,}\xspace}
\newcommand{\eg}{\emph{e.g.,}\xspace}
\newcommand{\etc}{etc.\xspace}
\newcommand{\etal}{et~al.\xspace}
\newcommand*{\tabref}[1]{\tablename~\ref{#1}\xspace}
\newcommand*{\figref}[1]{\figurename~\ref{#1}\xspace}
\newcommand*{\secref}[1]{Section~\ref{#1}\xspace}
\newcommand*{\eqref}[1]{Equation~\ref{#1}\xspace}

\newcommand*{\RQOne}{\newtext{How prevalent are SATDs in data-intensive systems?}}
\newcommand*{\RQTwo}{\newtext{How long do SATDs persist in data-intensive systems?}}

\newcommand*{\RQThree}{What is the composition of data-access SATD?}
\newcommand*{\RQFour}{What are the circumstances behind the introduction and removal of data-access SATD?}

\newcommand{\newtext}[1]{\textcolor{black}{#1}}
%\title{FIXME: Update Database Schema - An Empirical Study of Data Access Self-Admitted Technical Debt
\title{FIXME: Synchronize with Database}
\subtitle{An Empirical Study of Data Access Self-Admitted Technical Debt}


\author{Biruk Asmare 
Muse \and
        Csaba Nagy \and
        Anthony Cleve \and
        Foutse Khomh \and
        Giuliano Antoniol
}


\institute{Biruk Asmare Muse \at
              Polytechnique Montréal, \email{biruk-asmare.muse@polymtl.ca}
           \and
            Csaba Nagy \at
              Software Institute, Università della Svizzera italiana, \email{csaba.nagy@usi.ch}
             \and
            Anthony Cleve \at
              Namur Digital Institute, University of Namur, \email{anthony.cleve@unamur.be}
              \and
            Foutse Khomh \at
              Polytechnique Montréal, \email{foutse.khomh@polymtl.ca}
              \and
            Giuliano Antoniol \at
              Polytechnique Montréal, \email{antoniol@ieee.org}
}

\date{Received: date / Accepted: date}

\maketitle

\begin{abstract}

Developers sometimes choose design and implementation shortcuts due to the pressure from tight release schedules.
However, shortcuts introduce technical debt that increases as the software evolves.
The debt needs to be repaid as fast as possible to minimize its impact on software development and software quality.
Sometimes, technical debt is admitted by developers in comments and commit messages. Such debt is known as self-admitted technical debt (SATD).
In data-intensive systems, where data manipulation is a critical functionality, the presence of SATD in the data access logic could seriously harm performance and maintainability.
Understanding the composition and distribution of the SATDs across software systems and their evolution could provide insights into managing technical debt efficiently.
We present a large-scale empirical study on the prevalence, composition, and evolution of SATD in data-intensive systems.
We analyzed 83 open-source systems relying on relational databases as well as 19 systems relying on NoSQL databases.
We detected SATD in source code comments obtained from different snapshots of the subject systems.
To understand the evolution dynamics of SATDs, we conducted a survival analysis.
Next, we performed a manual analysis of 361 sample data-access SATDs, investigating the composition of data-access SATDs and the reasons behind their introduction and removal.
We identified 15 new SATD categories, out of which 11 are specific to database access operations. We found that most of the data-access SATDs are introduced in the later stages of change history rather than at the beginning. We also observed that bug fixing and refactoring are the main reasons behind the introduction of data-access SATDs.


\end{abstract}
% --------------------------------------------------------------------------
\section{Introduction}

With the increasing data demand of novel technologies, modern systems often collect and process large data volumes with high velocity for various purposes.
Such \textit{Big Data} or \textit{data-intensive systems} \cite{gokhale2008hardware} are pervasive and virtually affect people in all walks of life \cite{Park2021}.
They often have critical roles, too, calling for prime importance to ensure their quality.
Data-intensive systems, however, have several peculiarities posing challenges to software engineering practitioners and researchers \cite{CleveEtAl2010,Foidl2019,Hummel2018,Park2021}.

Developers of data-intensive systems are also often under pressure to deliver features on time.
Although deadlines can increase productivity, a potential adverse side effect is decreased quality \cite{Kuutila2020}.
This phenomenon led to the concept of \textit{technical debt}, \ie implementation trade-offs made by developers during rushed development tasks.
Since Cunningham first described technical debt almost 30 years ago \cite{Cunningham1992}, many researchers have studied its impact on software development \cite{Alfayez2020,Alves2016,Li2015,Rios2018}.
In general, researchers agree that it leads to low quality (in particular maintainability), and makes further changes more expensive in the long run \cite{Lim2012,wehaibi2016examining}.

Technical debt is often admitted by the developers through comments with ``todos'' and ``fixmes'' left in the source code as reminders for the future.
Such debt is referred to as \textit{self-admitted technical debt} (SATD).
Researchers often use SATD as a proxy to estimate technical debt because it can be identified by analyzing the source code \cite{liu2018satd,huang2018identifying} or issue reports \cite{Xavier2020}.

Technical debt is pertinent to data-intensive systems too.
In a recent study, Foidl \etal claim that technical debt can proliferate in data-intensive systems \cite{Foidl2019}.
As they say, data-intensive systems have heterogeneous architecture divided into multiple parts (software systems, data storage systems, and data), and debt introduced in one part has unwanted effects on other parts as well.
A similar phenomenon has been described by several authors \cite{LinNeamtiu2009,MeuriceEtAl2016,Stonebraker2017}, who found that changes in the database or application often remain unpropagated to the other side.
In the end, the system's quality decays over time \cite{Stonebraker2017}.

While these studies recognize the importance of technical debt in data-intensive systems, the problem has not received much attention.
Although several researchers have investigated the detection, persistence and impact of technical debt in traditional software systems (\eg \cite{Li2015,Alves2016,Rios2018,Alfayez2020,liu2018satd,Lim2012,wehaibi2016examining}), data-intensive systems remained out of focus.
In particular, we still do not know much about the prevalence and persistence of technical debt in data-intensive systems.
Neither do we know their composition and the circumstances of their introduction or removal.
This paper aims to fill this gap in the literature.

\newtext{
We conduct an empirical study to understand the prevalence and persistence of SATDs, their composition, and the circumstances of their introduction and removal. In particular, we seek answers to the following research questions.}
\begin{enumerate}[labelindent=\parindent,leftmargin=2\parindent]
    \item [\textit{RQ1:}] \textit{\RQOne} 
    \item [\textit{RQ2:}] \textit{\RQTwo}
    \item [\textit{RQ3:}] \textit{\RQThree}
    \item [\textit{RQ4:}]\textit{\RQFour}
\end{enumerate}

We focus on data-accesses and define \textit{data-access SATDs} as SATDs that occur in \textit{data-access classes}, \ie classes with direct database interactions or other persistence systems via calls to driver functions or APIs.
To differentiate against their counterparts, we refer to SATDs in non-data-access classes as \textit{regular SATDs}.
We are interested in data-access SATDs because mismanagement of SATDs in such classes can significantly impact the overall quality of a data-intensive system.

\newtext{
We examine SATDs in relational (SQL-based) and non-relational (NoSQL-based) data-intensive systems. The reason is the fundamental differences between SQL and NoSQL systems in terms of schema, data-access approach, data representation, scalability, and the type of data they are manipulating \cite{Sadalage2014,Scherzinger13,Scherzinger2020,Panos2021}. Relational systems have a predefined schema, use structured query language for data access, store data using tables, and capture the relationship between entities via the relationship between tables. They are vertically scalable and efficient for handling structured data. NoSQL systems have a dynamic schema, rely on document, key-value, graph, or column storage. They are horizontally scalable and efficient for handling unstructured data. SQL systems are often older and more mature compared to NoSQL systems. The conceptual differences in SQL and NoSQL systems result in differences in their APIs, affecting their data-access code, thus, the data-access SATDs too. Such differences motivated us to compare the prevalence and persistence of SQL-based and NoSQL-based data-intensive systems in our analysis. To the best of our knowledge, this is the first study on database-related technical debt that considers both relational and NoSQL software systems.
}

\noindent The primary contributions of this work can be summarized as follows.
\newtext{
\begin{enumerate}[nosep,labelindent=\parindent,leftmargin=2\parindent]
    \item We provide empirical evidence that SATDs are not equally prevalent between data-access and regular classes and between NoSQL and SQL systems.
    \item Our results show that data-access SATDs have lower survival than regular SATDs
    \item We extended the SATD taxonomy proposed by Bavota and Russo \cite{bavota2016large} with new SATD types, including 11 database access specific debts.
    \item Our result also shows that data-access SATDs are introduced at later stages of software evolution, mainly during \textit{bug fixing} and \textit{refactoring} activities.
\end{enumerate}
}

\noindent \textbf{The rest of the paper is organized as follows.}
We present related work in \secref{sec:lr}.
In \secref{sec:bg}, we provide background information on topic modeling and survival analysis.
We describe the study methodology in \secref{sec:m}, then we present the results in \secref{sec:sr}, and discuss their implications in \secref{sec:dis}.
In \secref{sec:ttv}, we identify the threats to the validity of our study.
Finally, we provide concluding remarks and discuss directions for future work in \secref{sec:conc}.

\section{Related work}
\label{sec:lr}

Several researchers have investigated self-admitted technical debt in source code for various purposes including its identification \cite{de2015contextualized,de2016investigating,da2017using,huang2018identifying,liu2018satd,yan2018automating,yu2020identifying,AlBarak2016}, removal \cite{zampetti2020automatically,maldonado2017empirical}, prioritization \cite{Alfayez2020,Albarak2018,kamei2016using}, recommendation when to admit SATDs \cite{zampetti2017recommending}, or the analysis of its impact on source code quality \cite{wehaibi2016examining} -- to mention a few examples. 

In this section, we first present an overview of recent literature and surveys related to technical debt. Then we summarize previous empirical studies on self-admitted technical debt. Finally, we discuss closely related papers focusing on technical debt in databases or data-intensive systems.

\subsection{Surveys and Literature Reviews on Technical Debt}

Li \etal conducted a systematic mapping study on technical debt and its management \cite{Li2015}. They examined 49 papers, classified technical debts into ten categories and identified eight activities and 29 technical debt management tools.  

Rios \etal performed a tertiary study and evaluated 13 secondary studies dating from 2012 to March 2018 \cite{Rios2018}. As a result, they developed a taxonomy of technical debt types and identified a list of situations in which debt items can be found in software projects.

Alves \etal performed a systematic mapping study by evaluating 100 studies dating from 2010 to 2014 \cite{Alves2016}. They also proposed a taxonomy of technical debt types and created a list of indicators to identify technical debt.

Alves \etal \cite{Alfayez2020} conducted a systematic literature review identifying 29 technical debt prioritization approaches. Among the 29 approaches, 70.83\% address a specific type of technical debt, while the remaining approaches can be applied to any kind of technical debt. 33.33\% of the approaches address code debt, 16.67\% address design debt, 12.5\% address defect debt and 1\% of the approaches is shared by SATDs, database normalization debt, requirement debt and architectural debt. Among all approaches, 54.17\% consider value and cost as prioritization decision factors, 29.17\%  rely on value only, and 16.67\% of approaches are based on value cost and constraint.  

Sierra \etal~\cite{sierra2019survey} present a survey of SATD studies from 2014 to 2019. They identified three main categories of research contributions: (1) papers that focus on the \emph{detection} of SATD, (2) papers that aim to improve the \emph{comprehension} of SATD, and (3) papers that focus on the \emph{repayment} of SATD.

Interestingly, while the above literature reviews identified various types of technical debt (\eg service debt related to web services), none of them  explicitly mention database communication-related debts in their taxonomies as well as evolution and management.
As they constitute an overview of the state-of-the-art technical debt research, the \textit{recent surveys indicate a lack of studies on database-related technical debt}. \newtext{We address this gap in the literature by extending the SATD taxonomy \cite{bavota2016large} with database access debt. Furthermore, we study the evolution and management of data-access SATDs to complement the state of the art.}

\subsection{Empirical Studies on Self-Admitted Technical Debt}

Potdar and Shihab~\cite{potdar2014exploratory} used source-code comments to study self-admitted technical debt in four large open-source software projects. They found that different types of self-admitted technical debts exist in up to 31\% of the studied project files. They showed that developers with higher experience tend to introduce most of the self-admitted technical debt and that time pressures and complexity of the code do not correlate with the amount of self-admitted technical debt introduced. They also observed that between 26.3\% and 63.5\% of self-admitted technical debt are removed from the projects after their introduction.

A large-scale empirical study on removing self-admitted technical debt was performed by Maldonado \etal, who examined 5,733 SATD removals in five large open source projects \cite{maldonado2017empirical}. They found that the majority (40.5–90.6\%) of SATD comments were removed from the systems, and the median amount of time that self-admitted technical debt stayed in the project ranged between 82–613.2 (18.2–172.8 days on average). 
\newtext{While the above studies address the prevalence and evolution of SATDs, they rely on four or five subject systems, limiting the generalization of the results, especially to data-intensive systems. We investigate the prevalence and evolution of data-access SATDs using 102 data-intensive subject systems.}

Bavota and Russo~\cite{bavota2016large} conducted a differentiated replication of the work of Potdar and Shihab~\cite{potdar2014exploratory}. They considered 159 software projects and investigated the diffusion (prevalence) and evolution of self-admitted technical debt and its relationship with software quality. Their results show that (1) SATD is diffused in software projects; (2) the most diffused SATDs are related to code, defect, and requirement debt; (3) the amount of SATD increases over time due to the introduction of new SATDs that developers never fix; and (4) SATD has very long survivability (over 1,000 commits on average). They also proposed a SATD taxonomy, which is used as a base for this work. \newtext{We extended their taxonomy by identifying data-access-specific SATDs.} 

\newtext {Wehaibi \etal \cite{wehaibi2016examining} studied the relation between SATD and software quality in terms of defects and maintenance effort. They identified SATDs in five popular open-source projects using pattern-based approaches. They found that the defectiveness of files increased after the introduction of SATDs and that changes were more difficult when they were related to SATDs.}

\newtext{Kamei \etal assessed ways to measure the interest of SATDs as a function of LOC and fan-in measures \cite{kamei2016using}. They examined JMeter as a case study and manually classified its SATD comments, then compared the metric values after the introduction and removal of SATDs to compute their interest. They found that up to 44\% of SATDs have positive interest implying that more effort is needed to resolve such debt.}

Zampetti \etal performed a quantitative and qualitative study of how developers address SATDs in five Java open source projects \cite{Zampetti2018}.
They found that a relatively large percentage (20\%–50\%) of SATD comments are accidentally removed while entire classes or methods are dropped.
Developers acknowledged in commit messages the SATD removal in only 8\% of the cases.
They also observed that SATD is often addressed by specific changes to method calls or conditionals, not just complex source code changes. \newtext{Like Zampetti \etal, we utilize the information obtained from commit messages to understand why data-access SATDs are introduced or removed.}

\newtext{
The work of Wehaibi \etal \cite{wehaibi2016examining} and Zampetti \etal \cite{Zampetti2018} motivated us to investigate the circumstances behind the introduction, evolution, and removal of data-access SATDs as such factors affect the interest of technical debt. We are interested in generalizing the findings of \cite{Zampetti2018} to the context of data-intensive systems.}

\newtext{\subsection{SATD detection approaches}}

\newtext{
Most of the SATD detection approaches are either pattern- or machine-learning-based. Initial methods for detection were pattern-based. Machine learning approaches were introduced more recently to improve the performance of detection approaches and tools.}
\newtext{
\subsubsection{Pattern-based SATD detection}
De Freitas \etal \cite{de2015contextualized} proposed a contextualized vocabulary model to identify technical debt using source code analysis. The model consists of software-related terms, adjectives that describe the terms, verbs to model actions in comments, adverbs, and tags such as Fixme and Todo. The combined terms can be used for searching comments in a pattern-based approach. They tested the feasibility of their approach on jEdit and Apache Lucene and identified technical debt in various categories.}

\newtext{
As an extension of the work of De Freitas \etal \cite{de2015contextualized}, De Farias \etal conducted an empirical study on the effectiveness of contextual vocabulary models (CVM) \cite{de2016investigating}. Besides evaluating the accuracy of the pattern-based approaches, they studied the impacts of language skills and developer experience on finding SATDs using a controlled experiment. Their result shows that the accuracy of the pattern-based approach looks promising, but it needs further improvement. English reading skills affected the identification of SATDs using pattern-based techniques.}

\newtext{\subsubsection{Machine-learning-based SATD detection}}
\newtext{
Maldonado \etal proposed NLP based approach to identify SATDs \cite {da2017using} automatically. Their approach can detect design and requirement SATDs. Furthermore, they built a manually labeled dataset of 62566 SATD comments. This dataset is used as a benchmark in most of the subsequent studies. They proposed a multi-class regression model using their dataset. They evaluated their approach and achieved an F1 measure between 40\% to 60\%. They observed that words related to sloppy code indicate design SATD while words related to incomplete code are associated with requirement SATD.}

\newtext{
Huang \etal proposed a machine-learning-based detection approach that combines the decisions of multiple Naive-Bayes-based classifiers into a composite classifier using majority vote \cite{huang2018identifying}. The comments from source codes are represented using vector space modeling (VSM), where features are selected utilizing Information Gain. 
They achieved an average F1-Score of 73.7\%.
%liu2018satd
Liu \etal \cite{liu2018satd} proposed a SATD detector tool which is a concrete implementation of Huang \etal \cite{huang2018identifying} approach. They provided this tool as a Java back-end library implementing the model to train and classify comments and the corresponding Eclipse plugin as a front end.} 
\newtext{
We also used this tool to detect SATDs in our subject systems due to its state-of-the-art detection performance and the availability of concrete implementation of the detection approach as a Java API and Eclipse plugin.
}

\newtext{Zampetti \etal \cite{zampetti2017recommending} presented TEDIOuS (TEchnical Debt IdentificatiOn System), a machine learning approach that provides recommendations to developers about ``technical debt to be admitted”. The method relies on source code structural metrics, readability metrics, and information from static analysis tools. They evaluated TEDIOus using nine open-source subject systems and achieved an average precision of 67\% and recall of 55\%. \\
}

\subsection{Technical Debt in Data-Intensive Systems}

Albarak and Bashoon defined the concept of database design debt as ``\textit{the immature or suboptimal database design decisions that lag behind the optimal/desirable ones, that manifest themselves into future structural or behavioral problems, making changes inevitable and more expensive to carry out}'' \cite{AlBarak2016}. They develop a taxonomy of debts related to the conceptual, logical, and physical design of a database. For example, they claim that ill-normalized databases (\ie databases with tables below the fourth normal form) can also be considered technical debt \cite{Albarak2018}. To tackle this specific type of debt, they propose an approach to prioritize tables that should be normalized.

Foidl \etal claim that technical debt can be incurred in different parts (\ie software systems, data storage systems, data) of data-intensive systems and different parts can further affect each other \cite{Foidl2019}. They propose a conceptual model to outline where technical debt can emerge in data-intensive systems by separating them into three parts: software systems, data storage systems and data. They present two smells as examples. Missing constraints, when referential integrity constraints are not declared in a database schema; and metadata as data, when an entity-attribute-value pattern is used to store metadata (attributes) as data. \newtext{While this study provided a conceptual model for components of data-intensive systems prone to technical debt, it did not provide empirical evidence for the existence of the technical debt. We contribute to addressing this gap by investigating SATDs in data-intensive systems.}

Weber \etal \cite{Weber2014} also identified relational database schemas as potential sources of technical debt. In particular, they provided a first attempt at utilizing the technical debt analogy for developing processes related to the missing implementation of implicit foreign key (FK) constraints. They discuss the detection of missing FKs, propose a measurement for the associated TD, and outline a process for reducing FK-related TD. As   illustrative case study, they consider OSCAR, a large Java medical record system used in Canada's primary health care.

Ramasubbu and Kemerer \cite{Ramasubbu2016} empirically analyze the impact of technical debt on system reliability by observing a 10-year life cycle of a commercial enterprise system. They also examine the relative effects of modular and architectural maintenance activities in clients. They conclude that technical debt decreases the reliability of enterprise systems. They also add that modular maintenance targeted to reduce technical debt is about 53\% more effective than architectural maintenance in reducing the probability of a system failure due to client errors.

\subsection{Summary}

The various studies and approaches discussed above constitute an extensive and sound basis for measuring, detecting and removing (self-admitted) technical debt. To the best of our knowledge, this paper is the first large-scale study investigating the prevalence, nature, and evolution of self-admitted technical debt in \emph{data-intensive} systems in general and in \emph{data-access} code in particular. It is also the first to study database-related technical debt in both relational and NoSQL software systems. In addition, it proposes an extension of an existing SATD taxonomy \cite{bavota2016large} to incorporate data-access related SATDs.

\section{Background}

\label{sec:bg}

This section provides a background on the topic modeling and survival analysis techniques used in our study.

\subsection{Topic Modelling}

\textit{Topic modeling} \cite{papadimitriou2000latent} is one of the unsupervised machine learning techniques that, given a set of documents (document corpus), can detect word and phrase patterns and cluster the documents based on word similarity. In our case, the corpus will be our dataset, and each comment will be one document in the corpus. Topic modeling works by counting the words and grouping documents with similar word patterns. Topic modeling is one of the frequently used techniques in \textit{natural language processing} (NLP). 

\textit{Latent semantic analysis} (LSA) \cite{papadimitriou2000latent} and \textit{latent Dirichlet allocation} (LDA) \cite{blei2003latent} are commonly used topic modeling algorithms. We also rely on LDA to assign topics to a set of words assuming that the arrangement of words determines the topic. LDA model is trained using a tokenized and pre-processed set of documents. After the LDA is trained, it can assign a document to a topic group with a certain probability. In this paper, we use LDA to cluster comments based on similarity so that our sampled data for manual analysis is not biased to a specific topic.
\newtext{LDA has hyper-parameters such as the number of topics, alpha, and beta to control the similarity levels that affect the model's performance. The first one determines the \textit{number of topics} generated by LDA after training. It can take any positive integer value. An insufficient value results in a too general model that makes topic interpretation difficult. An excess number of topics creates many topics that are too fine-grained for classification and subjective evaluation \cite{zhao2015heuristic}. \textit{Alpha} controls the document topic density. A higher alpha makes the documents contain many topics. On the contrary, a smaller alpha makes the documents have a small number of topics. \textit{Beta} controls the topic word density determining the number of words in the corpus associated with a topic. The higher the \textit{beta} value, the more words are associated with a topic. All those parameters need to be tuned using the target dataset by optimizing for the best performance of the LDA model.}

\newtext{\textit{Performance evaluation of LDA:}
A topic model can be evaluated by human judgment and intrinsic methods such as \textit{perplexity and coherence}. \textit{Perplexity} measures how well a probability model predicts a sample. It is computed by assessing the LDA model with unseen or held-out data. The lower the perplexity, the better the performance of the model. While perplexity measures the prediction of the LDA model, it does not evaluate the interpretation of the generated topics \cite{NIPS2009_f92586a2}. Another approach is to use coherence for evaluation. The coherence score is computed following segmentation, probability estimation, confirmation measure, and aggregation \cite{roder2015exploring}. Coherence score is calculated by summing the scores of a pair of words that describe a topic on the assumption that words that often appear together in the document are more coherent. Coherence takes a value between 0 and 1. The higher the score, the better the model.}

\subsection{Survival Analysis}
\label{sec:survival_analysis}

\textit{Survival analysis} \cite{miller2011survival} is a statistical analysis technique that provides the expected time for an event's occurrence. \textit{Time to event} and \textit{status} are two important variables for survival analysis. To compute each variable, we first need to define an event of interest that depends on the problem we want to analyze. In our case, an event of interest is the \textit{removal of an SATD}.

\textit{Time to event (T)} is defined as the time interval between the starting of observation (the first instance of the SATD) and the occurrence of an event of interest or the censoring of data. Time to event $T$ is a random variable with only positive values and can be measured in any unit \cite{miller2011survival}. The most common approach is to use time in minutes, hours, days, months, or years. However, we will use the number of commits to consider that the actual time may not correctly reflect software evolution compared to the number of commits. Projects have different activities at different times. Commits could be made more frequent at specific periods of time and less frequent at other times. Using time for $T$ in those cases has a limited capability to reflect project evolution. On the contrary, the \textit{number of commits} directly measures the project activity regardless of activity variation in some periods of time.

It is important to define an observation window and flag events outside it as censored. In our case, we define the observation window to cover all our snapshots of the subject systems. We flag SATDs that persist in the latest snapshots as censored since we do not know if the event of interest (\ie the removal of the SATD) will occur or not. Similarly, when an entire source file with one or more SATD comments is deleted within the observation window, we flag the SATDs as censored.
The reason is that in this case, it cannot be determined whether the SATDs are removed intentionally or only because of the file deletion. This is also supported by the observation of Zampetti \etal, who found that 20\%–50\% of the removals of SATDs are accidental and are even unintended \cite{Zampetti2018}.

Survival analysis takes a boolean variable called $\mathit{status}$ to distinguish between censored data and non-censored data. For instance, it takes a value of 1 when the event of interest occurred and 0 otherwise.

\textit{The survival function S(t)} gives the probability ($P(T > t)$) that a subject (SATD in our case) will survive beyond time $t$. 

After we computed $T$ and $\mathit{status}$, we can choose our survival estimator. We selected one of the commonly used survival estimators, the Kaplan-Meier estimator \cite{kaplan1958nonparametric}. The Kaplan-Meier estimation is computed following Equation \ref{eq:kapmai}. $t_{i}$ is the time duration (in the number of commits) up to event-occurrence (removal of SATD) point $i$, $d_{i}$ is the number of event occurrences up to $t_{i}$, and $n_{i} $ is the number of SATDs that survive just before $t_{i}$. $n_{i}$ and $d_{i}$ are obtained from the input data.

\begin{equation}
\label{eq:kapmai}
  S(t)= \prod_{i:t_{i} \leq t}{[1-\frac{d_{i}}{n_{i}}]}
\end{equation}

\subsection{\newtext{Metrics for measuring developers activity in time}}
\label{subsec:time-metrics}

Code repositories track changes in software artifacts through commits. The distribution of commits in time co-relates to developer activity and is used to study the evolution of software and the associated technical debts (\eg \cite{johannes2019large,tufano2017and}). For our analysis, we took a snapshot of projects every 500 commits. We provided the mean, standard deviation, and 95\% confidence interval of the commit time span for each SQL system\footnote{\url{https://bit.ly/2YLrLnU}} and NoSQL system\footnote{\url{https://bit.ly/3jj5JAH}} respectively in the replication package. 

\begin{figure}[!ht]
\centering
    \includegraphics[width=0.5\textwidth]{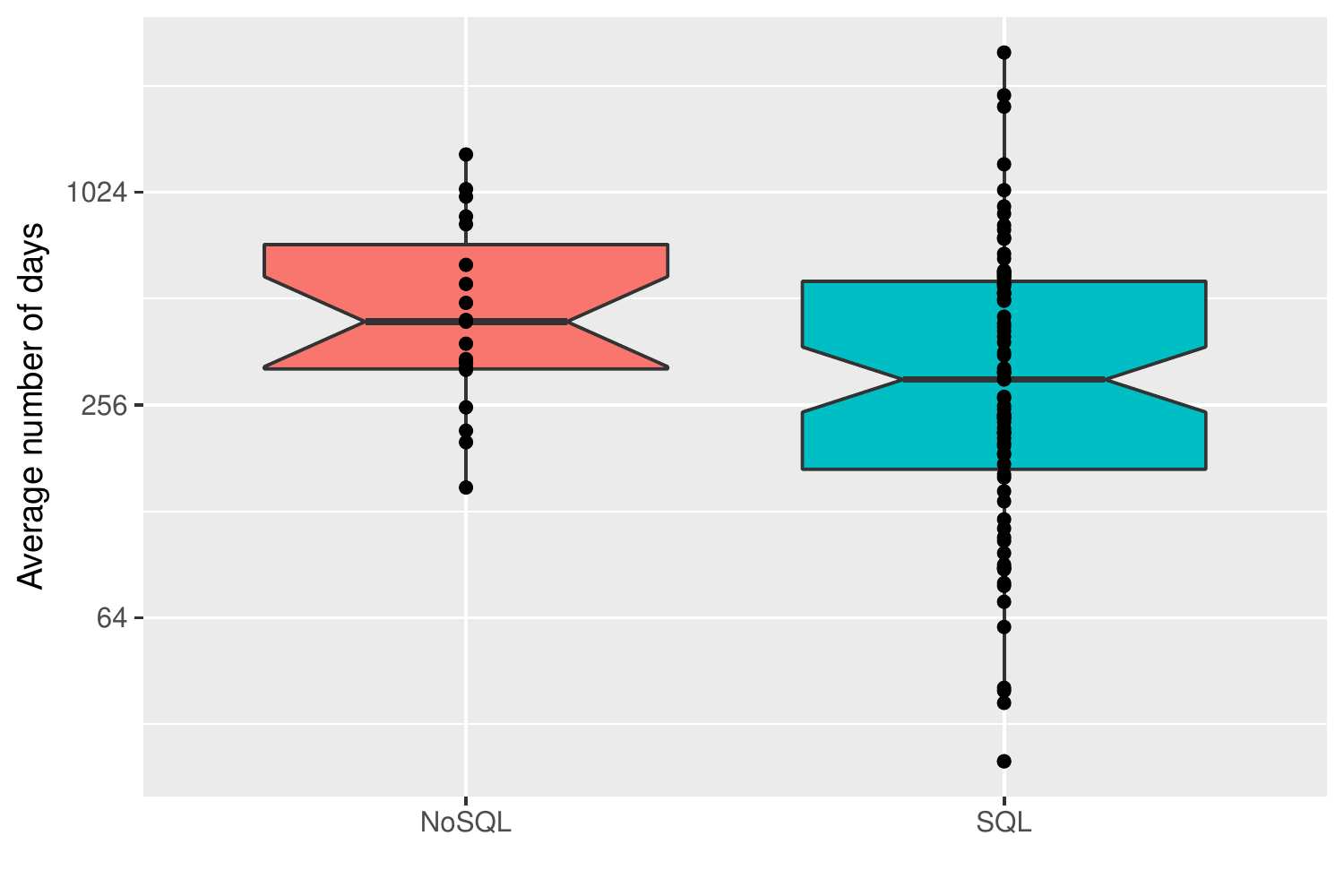}
    \caption{\newtext{The distribution of average time interval between successive snapshots taken every 500 commits for SQL and NoSQL subject systems. The y-axis time unit is in days.}}
    \label{fig:bgbp}
\end{figure}

Furthermore, Figure \ref{fig:bgbp} shows the distribution of the average time interval between successive snapshots of our subject systems. The average time interval between successive snapshots is 535 days for SQL subject systems and 423 days for NoSQL subject systems. The variation in time interval across and inside subject systems led to other approaches for measuring developer activity, such as using the number of commits.
While we use the number of commits to measure developer activity in our analysis, the above typical values can be used to interpret the commit time span in days.

\section{Study Method}

\label{sec:m}

\begin{figure*}[ht]
    \includegraphics[width=\textwidth]{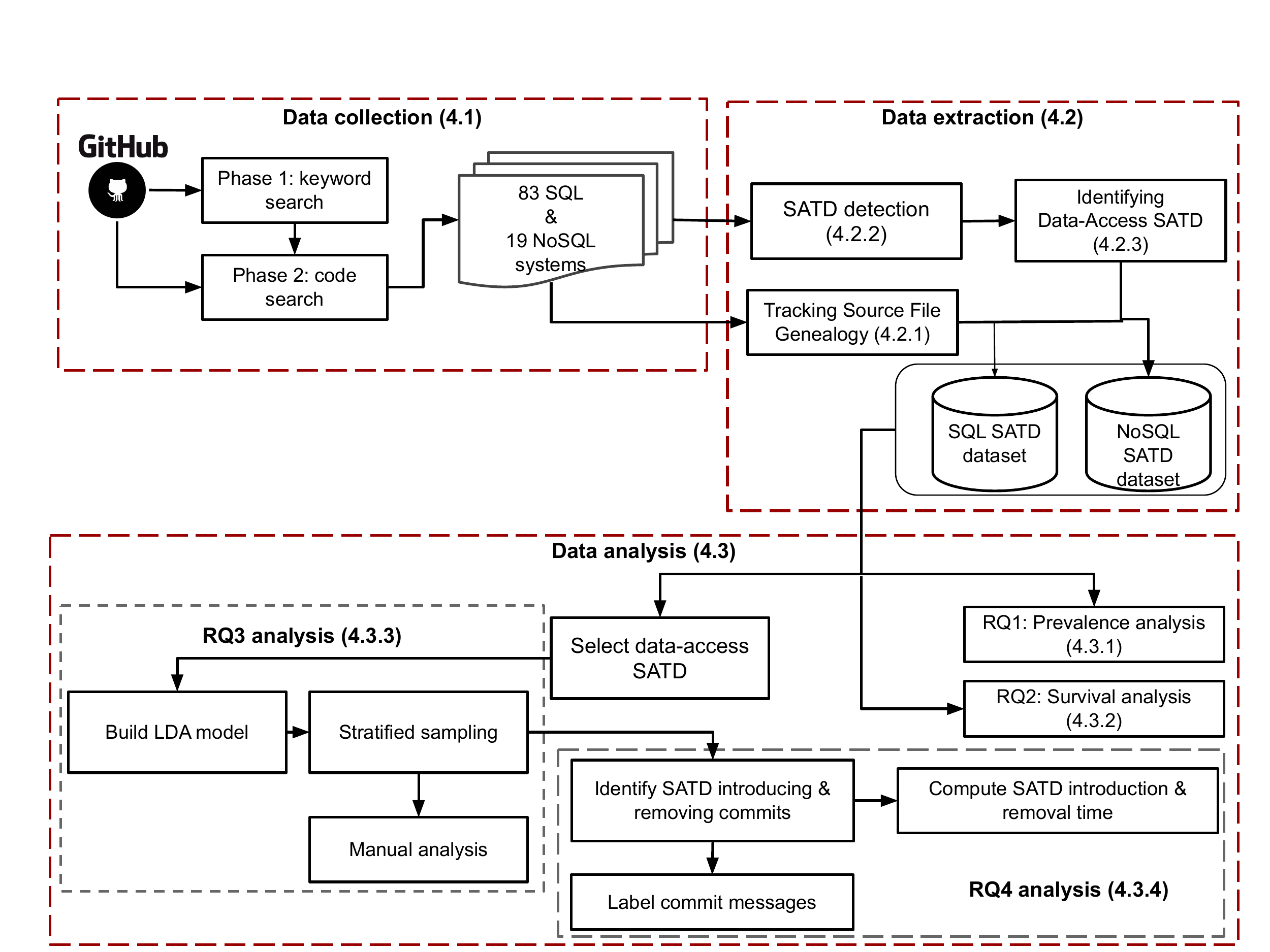}
     
    \caption{\newtext{Overview of the study method. We provided the subsection and sub-subsection numbers for easier matching with the description.}}
    
    \label{fig:Approach}
\end{figure*}

In this section, we discuss the approach we followed to answer our research questions. \figref{fig:Approach} gives an overview of our approach, including the subject system identification, data collection, and data analysis procedures. Each step is described in the coming sub-sections.

\subsection{Data collection}

We gathered SQL and NoSQL systems from GitHub for our study. We followed the following steps to identify subject systems.

\textbf{Phase 1}: We ran a GitHub search using the keywords related to the persistence libraries used by SQL projects such as SQLite, JDBC, Hibernate, and JPA. Those keywords are used assuming that projects that mention those libraries in the project name, description or readme file have a high chance of being data-intensive or having significant data-access code.

For NoSQL projects, we first collected NoSQL database management systems popular in open-source projects such as MongoDB, Redis, Riak and Neo4J.
The database systems are collected from the supported databases of Eclipse JNoSQL,\footnote{http://www.jnosql.org/docs/introduction.html} a popular Java framework in the Eclipse ecosystem that streamlines the integration of Java applications with NoSQL databases.
Currently, JNoSQL supports around 30 databases. The complete list of the databases is available in our replication package \cite{replication}.
We ran a GitHub search for projects mentioning these database engines.

To avoid ``toy'' projects, we set the following criteria for the projects: (1) they had to have at least one open issue, (2) more than 1,000 commits, and (3) at least one recent commit within one year from the time of data collection (\ie 2020).

\textbf{Phase 2}: We ran a code search on subject systems using the GitHub code search API \cite{github_code_search}.
We particularly looked for import statements for SQL projects corresponding to the persistence technologies such as Android SQLite API, JDBC, and Hibernate.
The import statements were identified in the work of Nagy \etal \cite{nagy2018sqlinspect}.
The import statements include, among others, \texttt{android.database.sqlite}, \texttt{android.database.DatabaseUtils}, \texttt{org.hibernate.Query}, \texttt{org. hibernate.SQLQuery},
\texttt{java.sql.Statement}, \texttt{javax.persistence.Query}, \texttt{javax.persistence. TypedQuery}, \texttt{org.springframework}.

Similarly, for NoSQL projects, we collected a list of import statements that are used to access NoSQL persistence systems, \eg \texttt{com.mongodb.MongoClient}, \texttt{org.neo4j.driver}, \texttt{org.apache. hbase}. To determine the imports, we started with the list of supported NoSQL databases from JNoSQL.
For each database, we explored code snippets provided as instructions to connect that database to Java applications.
We collected the import statements mentioned in such snippets to compile the list of NoSQL import statements.
We ran a code search on the identified projects and counted the import statements for each project.

We were finally left with 83 SQL and 19 NoSQL subject systems with the SQL and NoSQL import statements as mentioned above.
The complete list of the projects and import statements is available in our replication package \cite{replication}. 
\subsection{Data extraction}
\subsubsection{Tracking Source File Genealogy}

To ensure reliable evolution analysis, we need to keep track of all subject systems' source code genealogy.
We extracted file genealogy information using the \texttt{git diff} command. This command compares two snapshots and reports the added, modified, deleted and renamed files. Specifically, we used \texttt{git diff --name-status sha1 sha2}, where \texttt{sha1} and \texttt{sha2} are commit ids of the versions to be compared. \texttt{git diff} provides a numerical estimation of rename operations, which indicates the similarity percentage. In this work, renames with similarity thresholds greater than 70\% are considered true renames. A similar threshold was used in previous studies too \cite{johannes2019large,birukprev2019}. We tagged each source file with a unique ID generated from our file genealogy tracking database.   
\subsubsection{SATD Detection}

Due to the large number of subject systems with a large number of commits, we took snapshots of each system's every $500^{th}$ commits. Another approach would have been to select only a few projects and study every commit of each subject system. However, our research is exploratory and, therefore, we emphasize the generalizability of our results and conclusions.
A similar choice was made in other studies as well \cite{aniche2018code,birukprev2019}.

We used the SATD detection tool by Liu \etal \cite{liu2018satd}.
This tool uses a machine learning-based detection approach that combines the decisions of multiple Naive-Bayes classifiers into a composite classifier using a majority vote. During the tool's training phase, the source codes comments are represented using vector space modeling (VSM), and features are selected from VSM using information gain. The details of their approach are discussed in \cite{huang2018identifying}. The tool has a Java API as well as an Eclipse plugin to support developers. Given a source code comment, the tool returns a boolean indicating whether it is a SATD comment or not. We chose this detection tool because it has the highest accuracy (average $F_1$ score of 73.7\%) among different approaches, and the rest of the approaches were not realized as a tool to the best of our knowledge. 

SATD detection was carried out in two phases. In the first phase, we extracted the comments of each snapshot of all the projects using srcML.\footnote{https://www.srcml.org/} SrcML initially converts the source code into XML format. The comments can then be identified by running XPath queries. In the second phase, we run the SATD detection on the identified comments. The output of the SATD detection tool is binary: it classifies the comment as SATD-related or not. 

\subsubsection{Identifying Data-Access SATD}

We relied on SQL and NoSQL database import statements to identify data-access classes in both subject systems. We considered a class with at least one SQL/NoSQL database access import statement as a data-access class. To identify such classes, we ran a code search using the \texttt{egrep} command on all studied snapshots of the projects. An SATD comment that belongs to any of the identified data access classes is considered a data-access SATD.

\subsubsection{Study Dataset}

We built two SATD datasets corresponding to SQL and NoSQL subject systems. A row in each dataset contains \textit{fileId}, \textit{version}, \textit{commentId}, \textit{projectName}, \textit{commentMessage} and \textit{isDataAccess}. The \textit{version} attribute is needed because we study multiple versions of each subject system. Overall, our dataset contains 35,284 unique comments from SQL subject systems, out of which 4,580 are from data-access classes. Our dataset also contains 2,386 unique comments from NoSQL subject systems, out of which 436 are comments from data-access classes. 

\subsection{Data analysis}
\subsubsection{RQ1: \RQOne}

To answer RQ1, we computed the total number of data-access SATD comments and non-data-access SATDs for both SQL and NoSQL subject systems. We collected the number of SATDs for each snapshot of the subject systems' change history. We used violin plots to show how the prevalence of SATDs change as systems evolve and compared data-access and regular SATDs as well as SQL and NoSQL systems.

\subsubsection{RQ2: \RQTwo}

To answer this research question, we analyzed the persistence of SATDs using survival analysis. There are two cases when we automatically check if SATDs in File X are addressed between two versions A and B. Case 1: if an SATD comment in File X is similar between version A and B, we consider it as ``not fixed'' at version B. Case 2: if the comment found in version A is missing in version B, we consider it ``fixed'' at version B. We choose the number of commits over time in days for the survival analysis because different projects have different activities in time. As we discussed before (see \secref{sec:survival_analysis}), the number of commits suits better than time for our purpose to reflect the projects' activity \cite{bavota2016large}.

We used the Kaplan-Meir curve to visualize the survival of subject SATDs. The Kaplan-Meir curve shows the survival probability $S(t)$ of a given SATD at a time $t$. We define the addressing of a SATD as our event of interest. The occurrence of this event determines the survival probability. SATDs that persisted up to the latest versions and those removed with the source files are flagged as censored (see \secref{sec:survival_analysis}). 

\subsubsection{RQ3: \RQThree}
\newtext{To answer this research question, we first identified unique data-access SATD comments in our dataset. We built an LDA topic model on the dataset to generate the strata needed for stratified sampling. Finally, We conducted a manual analysis on the sample SATD comments. We provide a detailed description in the following paragraphs.}

\newtext{\textit{Build LDA model:} } We then applied common NLP preprocessing techniques. In particular, we removed punctuation, common English stop words, and the words ``todo'' and ``fixme,'' as they are very common in most comments.

Then, we applied lemmatization and stemming using the Python NLTK library.
The final output of this pre-processing is a tokenized comment. 

 The tokenized comments were transformed using TF-IDF transformation. The input of the LDA was the TF-IDF representation of the comments in our dataset.
 After the preprocessing, we run the LDA topic model, using the Gensim Python library to cluster the SATD comments based on similarity. We experimented with different hyper-parameters, \newtext{namely the number of topics, alpha, and beta using coherence score as model performance evaluation. First, we experimented with topics from 5 to 75, increasing by 5 every iteration. The coherence score gradually increased as we increased the number of topics and reached a maximum value of 20 topics (0.39\%) for SQL systems. For NoSQL systems, we started with less than five topics since the corpus of NoSQL comments was smaller, then we continued until 150 because we saw some fluctuations in coherence score as the number of topics increased. We obtained the highest coherence score (0.45\%) when the number of topics was set to 4. Next, we experimented with alpha and beta using a range from 0.01 to 0.1 with 0.3 intervals. We did not see a significant change in the coherence score. Hence, we used the default values on Gensim (alpha and beta: `symmetric' meaning alpha and beta are set as the inverse of the number of topics). Both LDA models achieved a lower coherence score below 0.5. However, we did not consider the interpretation of the topics. Instead, we used the LDA to cluster similar comments before sampling}. After the LDA model training, we assigned each document to a specific topic. The overall output of the LDA model was documents clustered under each topic group. We used stratified random sampling from the clusters to generate our dataset for manual analysis.

 \newtext{\textit{Stratified sampling:}} We prepared a dataset for manual analysis using stratified sampling from each LDA topic group. The dataset contains 183 data-access SATD comments for SQL systems and 178 data-access SATD comments for NoSQL systems. We used stratified random sampling to pick representative samples from each LDA topic group or cluster.

\newtext{\textit{Manual analysis:} }
The manual analysis was conducted using deductive coding to assign themes to the comments. We started with the themes identified by Bavota and Russo \cite{bavota2016large} and extended them with themes specific to data access. To have a common interpretation of the labels among authors, we conducted iterative sessions to label sample SATDs and resolve conflicts. After that, the first author labeled all the 361 SATDs, which were then divided into three sets and reviewed by three more authors to ensure that at least one additional person checks each label. Finally, all conflicts were resolved through face-to-face discussions.

During the labeling process, we found some comments that were identified as SATD comments by the detector tool but were not related to technical debt. Recall that Liu \etal reported an average $F_1$ score of 73.7\% for the tool \cite{liu2018satd}. A common reason was that they contained one of the keywords (\eg ``fix''), but the developer meant it for a different purpose (\eg ``\textit{// import release fix into the release branch}''\footnote{https://bit.ly/3siSWzX}). We found 105 instances (29\%) of these comments and marked them as \textit{false positives}.

We found 12 comments in which the information from the comments and source code did not give enough context to assign the comments to the appropriate category. We marked such instances as \textit{unclear}.

We also found 4 comments that belonged to more than one category as they typically ordered tasks in a list under a ``todo'' comment. These tasks often belonged to various SATD categories; hence, we decided to mark them as \textit{multi-label comments}.

For multi-label comments, we cannot identify one specific category. Hence we exclude them for RQ3 but keep them for the evolution-related research question RQ4. After we excluded false positives, unclear comments and multi-label comments, the final dataset contained a total number of 240 data-access SATD comments.

\subsubsection{RQ4: \RQFour}

\newtext{In our analysis, we use the introduction or removal of SATD comments as a proxy to the introduction or removal of SATDs, respectively. We are particularly interested in investigating when and why the data-access SATDs are introduced and removed. Hence, we first identified the SATD introduction and removal commits and then computed the commit time. Then we conducted manual labeling of the commit messages. We outline the details of our analysis in the following paragraph.}

\newtext{\textit{Identify SATD introducing and removing commits:}}  Using this labeled data from RQ3, we extracted the commits that introduced the comments and commits that removed them from the change history of the subject systems. We used the PyDriller repository mining framework \cite{PyDriller} for our analysis. PyDriller is used to analyze both local and remote repositories and extract information related to their change history. We looked for the \textit{SATD introducing commit} given the path of a file by looking at the change history starting from the beginning to the end and looking for the first occurrence of the SATD under study. Similarly, we looked for the \textit{SATD removal commit}, the commit in which a SATD is removed from the system, by looking for the first commit in which the SATD is no longer present given that the SATD occurred in the previous versions. To check if the SATD is removed together with the hosting class, we also keep track of the commit where the hosting class is removed (if it is removed).

\paragraph{When are data-access SATDs introduced or removed?}

For our purpose, the number of commits is better than the absolute time at reflecting software evolution (see Section~\ref{sec:survival_analysis}).
Hence, we measure introduction time and removal time in terms of number of commits.

We define \textit{introduction time} ($t_i'$) as the number of commits that occurred before and including the first occurrence of the SATD under study.

Similarly, we define \textit{removal time} ($t_r'$) as the number of commits that occurred before and including the commit that removed the SATD.
$t_i'$ and $t_r'$ are measured in the number of commits.

Since the total number of commits varies across the projects, we normalize the \textit{introduction time} and \textit{removal time} with the total number of commits for each subject system (see Equations \ref{eq:introtime} and \ref{eq:removaltime}). We use a similar normalization for the removal time. For example, a SATD introduction time of 20\% for a project with 1,000 commits means the SATD was introduced in the 200$^{th}$ commit from the beginning. The smaller the value, the closer the introduction of SATD to the early stages of the project evolution and vice versa.

\begin{equation}
\label{eq:introtime}
  \mathit{Introduction\, time} = \frac{t_i' \cdot 100}{\mathit{Total\, number\, of\, commits}}, 
\end{equation}

\begin{equation}
\label{eq:removaltime}
  \mathit{Removal\, time} = \frac{t_r' \cdot 100}{\mathit{Total\, number\, of\, commits}}
\end{equation}

We use \textit{Introduction time} and \textit{Removal time} to investigate when SATDs are introduced or removed.

\paragraph{Why are data-access SATDs introduced or removed?}

To investigate why data-access SATDs are introduced, we \newtext{collected the commit messages of SATD introduction and removal commits, then manually categorized their goal or purpose}. We use similar categories to Tufano \etal~\cite{tufano2015and}: \textit{bug fixing}, \textit{enhancement}, \textit{new feature}, and \textit{refactoring}. In our case, we added \textit{merging} and \textit{multiple goals} to account for merging commits and commits whose messages have more than one goal.
In this way, the commit goal can be mapped to more than one of the categories from Tufano \etal 

\textit{Bug fixing} commits mention that the commit was made to fix an existing bug or issue.
\textit{Enhancement} commits aim at enhancing existing or already implemented features.
Commits with the goal \textit{new feature} describe their goal as introducing or supporting a new functionality.
Commits that mention refactoring operations are categorized under \textit{refactoring}.
Finally, commits made for merging pull requests and branches are categorized under \textit{merging}.
We labeled SATD-removing commits similarly.

\section{Study Result}
\label{sec:sr}

In this section, we present the results of our study. The raw results and corresponding analysis are reported for each research question.

\subsection{RQ1: \RQOne}

In this subsection, we present the prevalence of SATD in SQL and NoSQL systems.

We also show how data-access and regular SATDs evolve in multiple snapshots of the systems. 

\begin{figure}[!ht]
\centering
    \includegraphics[width=0.5\textwidth]{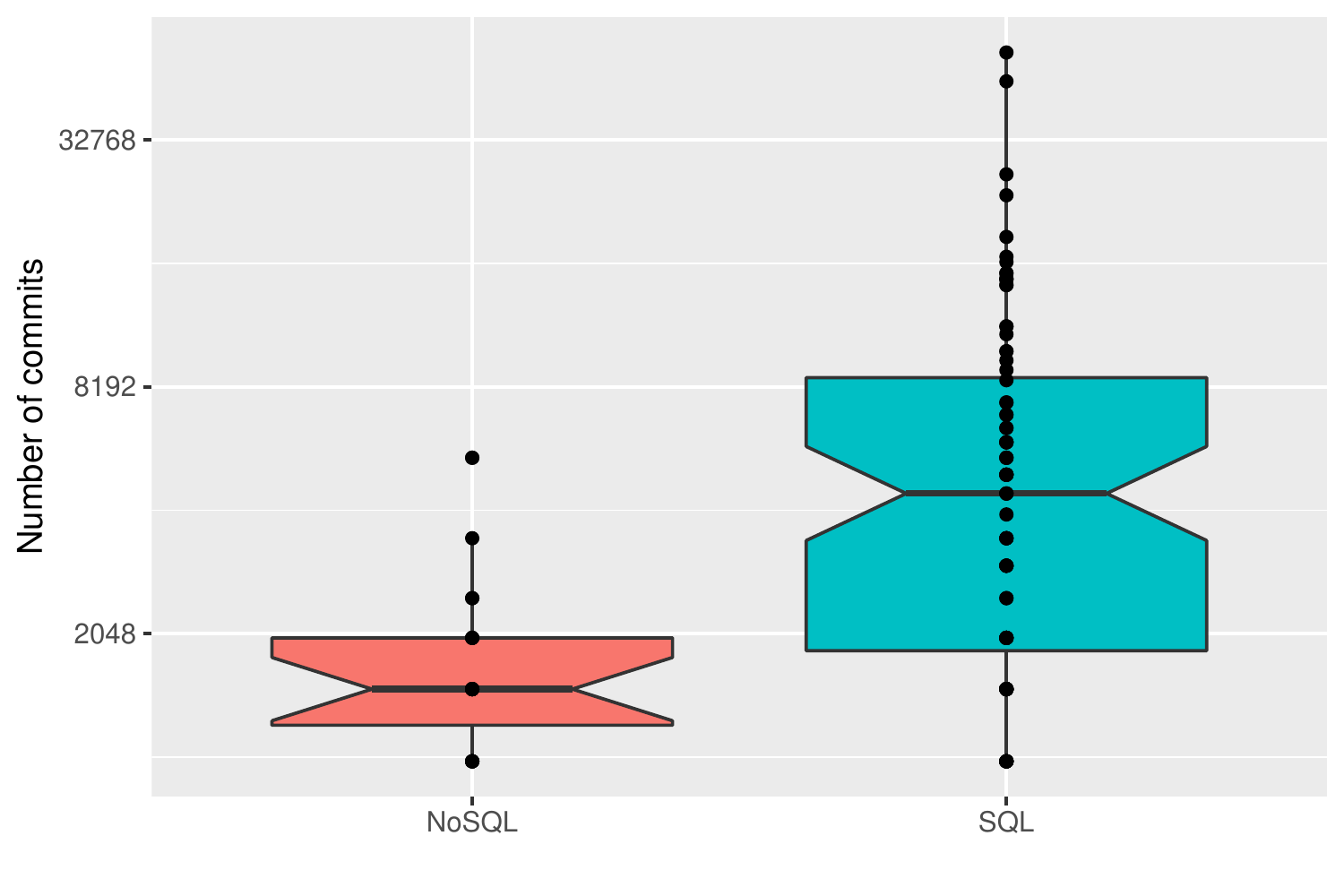}
    \caption{\newtext{Distribution of the number of commits in SQL and NoSQL subject systems. The y-axis is on a log scale.}}
    \label{fig:rq1projs}
\end{figure}

\figref{fig:rq1projs} shows the distribution of the number of commits for SQL and NoSQL systems.

We can see a significant difference in the number of commits between the two types of systems.
SQL systems have a median of 4,501 and a mean of 7,066.5 commits.
The maximum number of commits is 53,501 for SQL subject systems.
On the other hand, for NoSQL systems, the median number of commits is 1,501, and the mean is 1,869.42. The maximum number of commits is 5,501 for NoSQL systems.

\begin{table}[!ht]
  \centering
  \caption{Project groups}
  \begin{tabular}{lrrrr} \toprule
      \textbf{Group} & \textbf{Min. Commits} & \textbf{Max. Commits} & \textbf{NoSQL projects} & \textbf{SQL projects} \\
      \midrule
      $Group_1$ & 1001    & 1,500  & 12 & 21 \\
      $Group_2$ & 1,501 & 6,750  & 7 & 37 \\
      $Group_3$ & 6,751 & 53,501 & 0 & 26 \\
     \bottomrule
  \end{tabular}
  \label{tab:groups_profile}
\end{table}

 The quantile analysis of the distribution of commits shows that 25\% of the projects have less than 1,501 commits, 50\% of the projects less than 3,001, and 75\% of the projects less than 6,751 commits. \newtext{We grouped the projects based on the quantiles into three for the purpose of visualization.} 
\tabref{tab:groups_profile} presents a summary of the systems in each group.
For example, all projects with a maximum of 1,500 commits are included in $Group_1$, including 12 NoSQL and 21 SQL subject systems.

\begin{table}[]
\centering
\newtext{
\caption{Summary of the distribution of data-access and regular SATDs over the number of commits in Group 1 subject systems}
\label{tbl:rq1-g1-dist}
\resizebox{\textwidth}{!}{
\begin{tabular}{@{}ll|lcrccr|lcrccr@{}}
\toprule
 &  & \multicolumn{6}{c}{\textbf{Data-access SATD}} & \multicolumn{6}{c}{\textbf{Regular SATD}} \\ \midrule
Commit & System & Min & 25\% & Mean & \textbf{Median} & 75\% & Max & Min & 25\% & Mean & \textbf{Median} & 75\% & Max \\ \midrule
1 & NoSQL & 0 & 0 & 1.92 & \textbf{0} & 0.25 & 19 & 1 & 8.25 & 47.83 & \textbf{23.5} & 47.25 & 304 \\
 & SQL & 0 & 0 & 5.05 & \textbf{0} & 3.5 & 31 & 1 & 8 & 35.26 & \textbf{12} & 31.5 & 281 \\ \midrule
501 & NoSQL & 0 & 0 & 5.75 & \textbf{1} & 8.5 & 24 & 0 & 7.5 & 79.67 & \textbf{38.5} & 87 & 477 \\
 & SQL & 0 & 1 & 17.20 & \textbf{1.5} & 8.25 & 163 & 1 & 13.75 & 57.75 & \textbf{28.5} & 64.25 & 412 \\ \midrule
1001 & NoSQL & 0 & 2 & 13.50 & \textbf{4} & 14.5 & 64 & 2 & 6 & 74.42 & \textbf{35} & 95.5 & 370 \\
 & SQL & 0 & 1 & 27.76 & \textbf{4} & 17 & 226 & 1 & 13 & 63.67 & \textbf{35} & 74 & 293 \\ \midrule
1501 & NoSQL & 1 & 2 & 34.71 & \textbf{22} & 43.5 & 129 & 4 & 17 & 96.57 & \textbf{71} & 88 & 391 \\
 & SQL & 1 & 2.5 & 63.17 & \textbf{5.5} & 75.25 & 380 & 12 & 20.5 & 73.67 & \textbf{40} & 84.5 & 290 \\ \bottomrule
\end{tabular}
}
}
\end{table}

\begin{table}[]
\centering
\caption{Summary of the distribution of data-access and regular SATDs over the number of commits in Group 2 subject systems}
\label{tbl:rq1-g2-dist}
\newtext{
\resizebox{\textwidth}{!}{
\begin{tabular}{@{}ll|lcrccr|lcrccr@{}}
\toprule
 &  & \multicolumn{6}{c}{\textbf{Data-access SATD}} & \multicolumn{6}{c}{\textbf{Regular SATD}} \\ \midrule
Commit & System & Min & 25\% & Mean & \textbf{Median} & 75\% & Max & Min & 25\% & Mean & \textbf{Median} & 75\% & Max \\ \midrule
1 & NoSQL & 0 & 0 & 0.57 & \textbf{0} & 0 & 4 & 4 & 5 & 21.14 & \textbf{10} & 29 & 66 \\
 & SQL & 0 & 0 & 6.89 & \textbf{0} & 0.25 & 203 & 1 & 7 & 64.61 & \textbf{24} & 86.25 & 580 \\ \midrule
1001 & NoSQL & 0 & 0 & 7.00 & \textbf{4} & 12 & 21 & 2 & 7.5 & 30.86 & \textbf{15} & 46.5 & 91 \\
 & SQL & 0 & 0 & 8.14 & \textbf{0} & 2 & 121 & 3 & 18 & 111.16 & \textbf{40} & 164 & 586 \\ \midrule
2001 & NoSQL & 0 & 3 & 13.14 & \textbf{7} & 11.5 & 56 & 0 & 15.5 & 35.43 & \textbf{26} & 50 & 91 \\
 & SQL & 0 & 0 & 19.03 & \textbf{2} & 6 & 316 & 5 & 29 & 147.76 & \textbf{51} & 239 & 1015 \\ \midrule
3001 & NoSQL & 1 & 5.5 & 10.00 & \textbf{10} & 14.5 & 19 & 17 & 25.5 & 34.00 & \textbf{34} & 42.5 & 51 \\
 & SQL & 0 & 1 & 31.30 & \textbf{5} & 9 & 506 & 4 & 37 & 160.37 & \textbf{87} & 224.5 & 857 \\ \midrule
4001 & NoSQL & 2 & 2 & 2.00 & \textbf{2} & 2 & 2 & 20 & 20 & 20.00 & \textbf{20} & 20 & 20 \\
 & SQL & 0 & 1 & 40.17 & \textbf{2.5} & 10.5 & 555 & 24 & 40.75 & 169.83 & \textbf{85} & 220.25 & 923 \\ \midrule
5001 & NoSQL & 18 & 18 & 18.00 & \textbf{18} & 18 & 18 & 11 & 11 & 11.00 & \textbf{11} & 11 & 11 \\
 & SQL & 0 & 1.5 & 54.13 & \textbf{4} & 12 & 588 & 3 & 39.5 & 179.60 & \textbf{95} & 266 & 941 \\ \midrule
6001 & SQL & 1 & 1.25 & 26.67 & \textbf{2.5} & 3 & 150 & 11 & 27.5 & 48.67 & \textbf{33.5} & 76.25 & 98 \\ \bottomrule
\end{tabular}
}
}
\end{table}

\begin{table}[]
\centering
\caption{Summary of the distribution of data-access and regular SATDs over the number of commits in Group 3 SQL subject systems}
\label{tbl:rq1-g3-dist}
\newtext{
\resizebox{\textwidth}{!}{
\begin{tabular}{@{}l|lcrccr|lcrccr@{}}
\toprule
 & \multicolumn{6}{c}{\textbf{Data-access SATD}} & \multicolumn{6}{c}{\textbf{Regular SATD}} \\ \midrule
Commit & Min & 25\% & Mean & \textbf{Median} & 75\% & Max & Min & 25\% & Mean & \textbf{Median} & 75\% & Max \\ \midrule
1 & 0 & 0 & 3.46 & \textbf{0} & 0 & 60 & 4 & 27.5 & 203.96 & \textbf{67.5} & 153.75 & 1485 \\
10001 & 0 & 1 & 71.00 & \textbf{18} & 50.5 & 519 & 99 & 203 & 552.47 & \textbf{307} & 648.5 & 2263 \\
20001 & 0 & 2.5 & 10.00 & \textbf{5} & 15 & 25 & 177 & 183 & 189.33 & \textbf{189} & 195.5 & 202 \\
30001 & 0 & 0.75 & 1.50 & \textbf{1.5} & 2.25 & 3 & 180 & 192.75 & 205.50 & \textbf{205.5} & 218.25 & 231 \\
40001 & 0 & 0.75 & 1.50 & \textbf{1.5} & 2.25 & 3 & 202 & 223.75 & 245.50 & \textbf{245.5} & 267.25 & 289 \\
50001 & 4 & 4 & 4.00 & \textbf{4} & 4 & 4 & 308 & 308 & 308.00 & \textbf{308} & 308 & 308 \\ \bottomrule
\end{tabular}
}
}
\end{table}

\begin{figure}[!ht]
    \centering
    \subfloat[Regular SATD\label{fig:rq1ndag1}]{%
        \centering
        \includegraphics[width=0.49\textwidth]{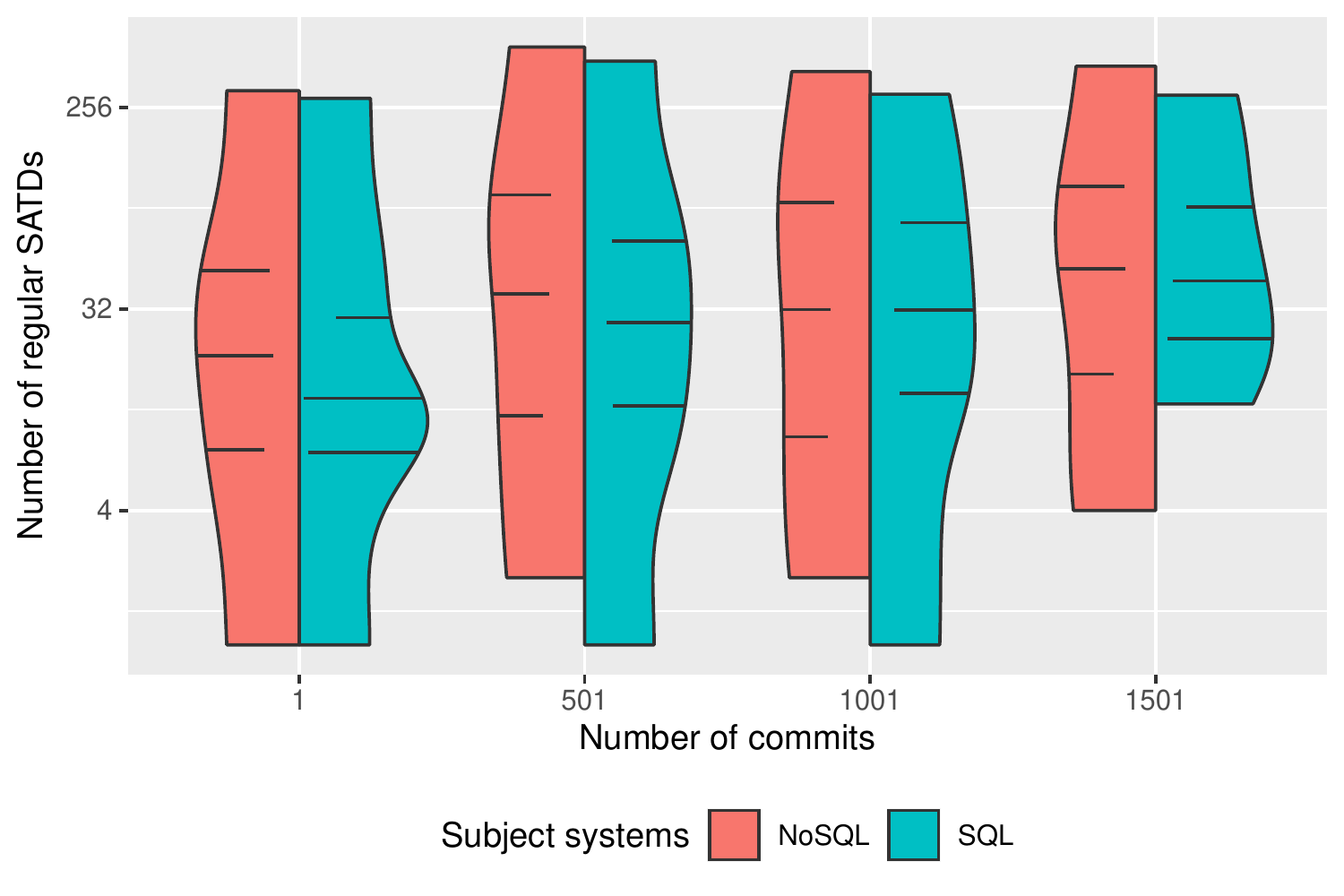}
    }\hfill
    \subfloat[Data-access SATD\label{fig:rq1dag1}]{%
        \centering
        \includegraphics[width=0.49\textwidth]{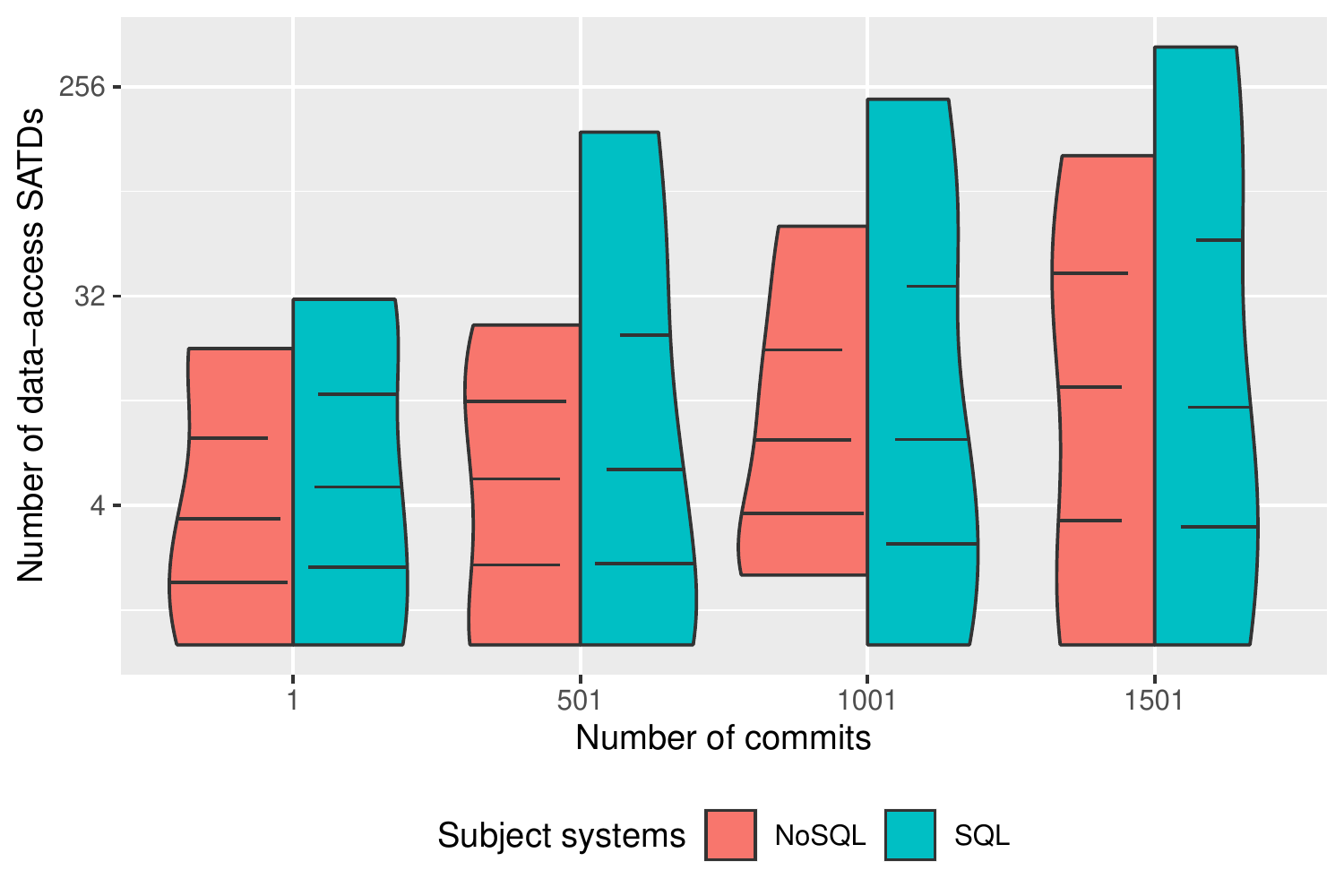}
    }
    \caption{\newtext{Prevalence of regular and data-access SATD in $Group_1$. The horizontal lines in this and subsequent violin plots show the 25\%, median, and 75\% quantiles respectively from bottom to top.}}
\end{figure}

\newtext{Tables \ref{tbl:rq1-g1-dist}, \ref{tbl:rq1-g2-dist} and \ref{tbl:rq1-g3-dist} show the summary of the distribution of SATDs in our subject systems by the project groups. The distribution was computed over the snapshots of the subject systems. }
\figref{fig:rq1ndag1} shows the distribution of regular SATDs in $Group_1$.
We observe that the number of regular SATDs increases for SQL systems as the number of commits increases.
For NoSQL systems, an increase in the SATDs is observed, moving from 1 to 501 and 1,001 to 1,501. 
The median of regular SATDs in NoSQL systems (23.5, 38.5, 71) is higher than in SQL systems (12, 28.5, 40) for snapshots at commits 1 and 501 and 1,501, respectively. The highest number of regular SATDs (477) was observed at the $501^{st}$ commit of a NoSQL system, \textit{Bboss}.\footnote{https://github.com/bbossgroups/bboss}
Bboss is a framework that provides API support for developing enterprise and mobile applications.

\figref{fig:rq1dag1} shows the distribution of data-access SATDs in $Group_1$.
The number of data-access SATDs in $Group_1$ increases with the number of commits.
The median data-access SATD for SQL systems is 0, 1.5, 4, and 5.5 for commits 1, 501, 1,001 and 1,501.
For NoSQL systems, the median is 0, 1, 4, and 22 for commits 1, 501, 1,001 and 1,501, respectively.
We can see that the median of data-access SATDs is roughly similar between SQL and NoSQL subject systems except for commit 1,501, where we observe a large difference in magnitude between SQL and NoSQL subject systems.
The highest number of data-access SATDs (380) was observed at commit 1,501 by the SQL subject system \textit{Blaze-persistence}.\footnote{https://github.com/Blazebit/blaze-persistence}
Blaze-persistence is a criteria API provider project for applications that rely on JPA for data persistence. 

\begin{figure}[!ht]
    \centering
    \subfloat[Regular SATD\label{fig:rq1ndag2}]{%
        \centering
        \includegraphics[width=0.49\textwidth]{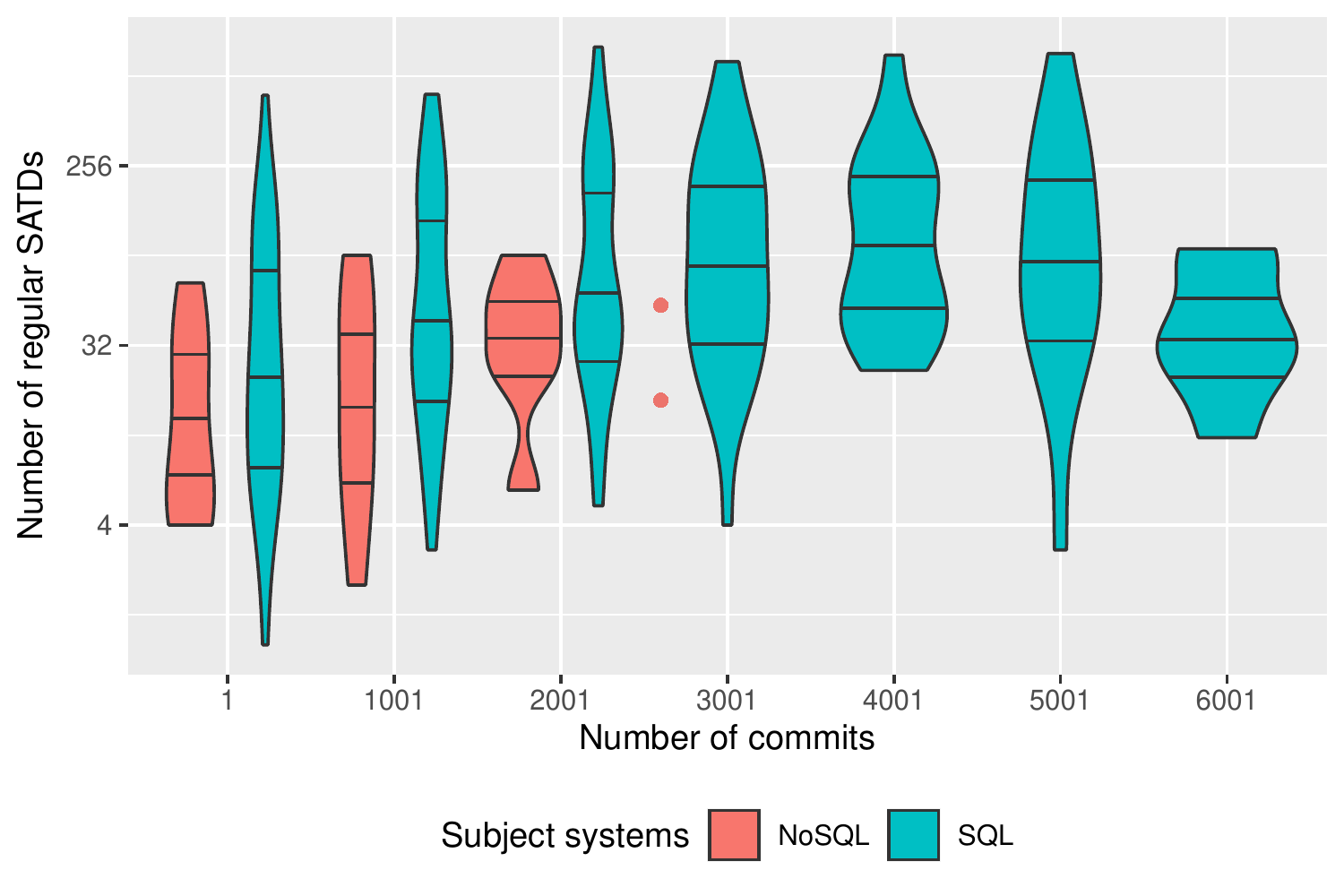}
    }\hfill
    \subfloat[Data-access SATD\label{fig:rq1dag2}]{%
        \centering
        \includegraphics[width=0.49\textwidth]{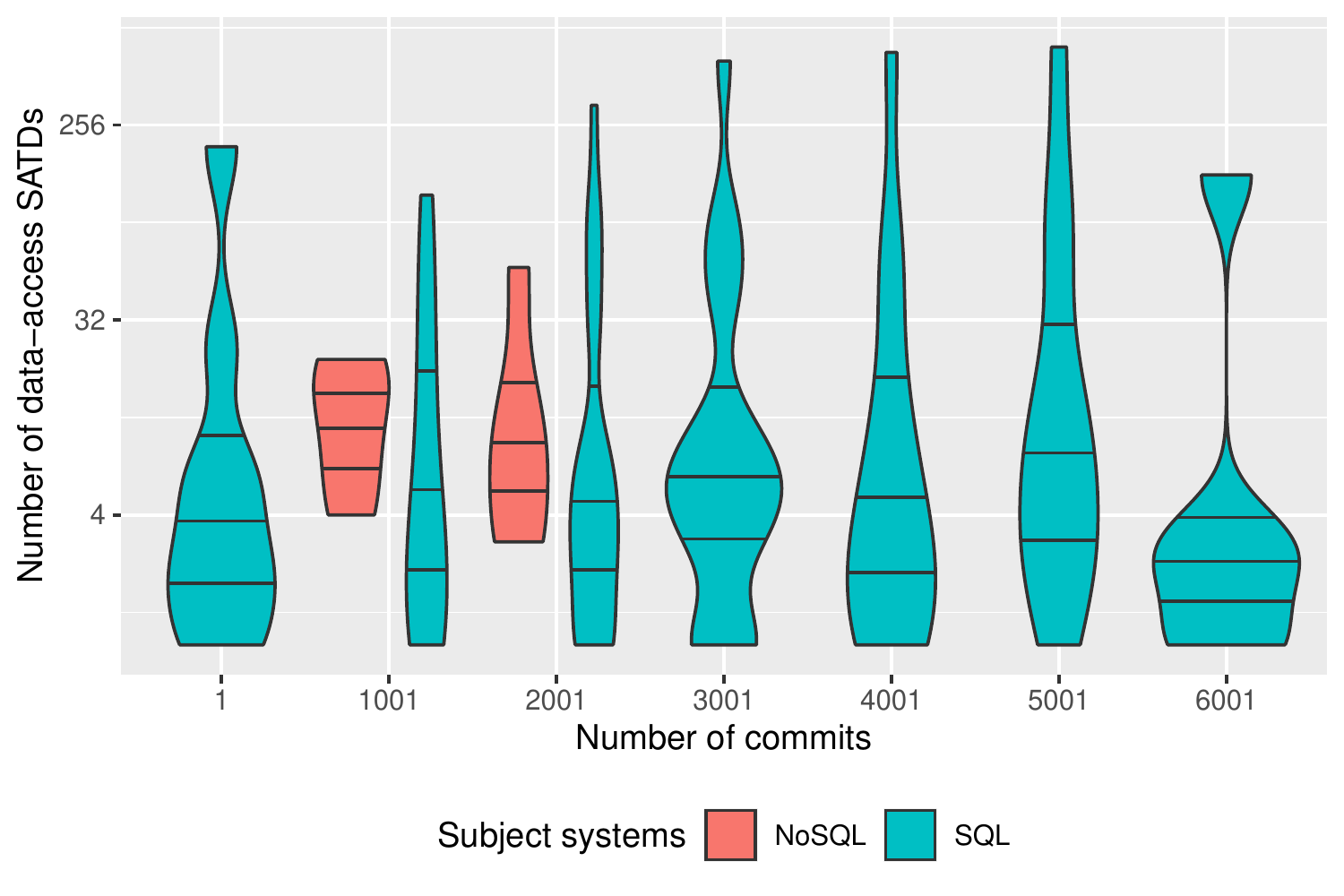}
    }
    \caption{Prevalence of regular and data-access SATD in $Group_2$.}
\end{figure}

\figref{fig:rq1ndag2} shows the distribution of regular SATDs in $Group_2$.
We observe an increasing trend in the number of regular SATDs for both SQL and NoSQL systems.
SQL systems have a higher median number of regular SATD in all snapshots.
The median number of regular SATD of SQL systems is 24, 40, 51 and 87 for commits 1, 1,001, 2,001 and 3,001, respectively.
For NoSQL systems, the median is 10, 15, 26 and 34, respectively.
The maximum number of regular SATD (1,015) was registered in an SQL system, \textit{Jena-sparql-api},\footnote{https://github.com/SmartDataAnalytics/jena-sparql-api} at commit 2,001.
\textit{Jena-sparql-api} provides a SPARQL processing stack for building Semantic Web applications. 

We observe a similar trend of increase in the number of data-access SATDs on $Group_2$, as shown in \figref{fig:rq1dag2}.
The median number of data-access SATD in NoSQL systems is 0, 4, and 7 for commits 1, 1,001, and 2,001.
SQL systems have a median number of data-access SATD 0, 0, and 2, respectively.
The largest data-access SATD (588) was registered at commit 5,001 by SQL system \textit{Threadfix},\footnote{https://github.com/denimgroup/threadfix} a software vulnerability management application. 

\begin{figure}[!ht]
    \centering
    \subfloat[Regular SATD\label{fig:rq1ndag3}]{%
        \centering
        \includegraphics[width=0.49\textwidth]{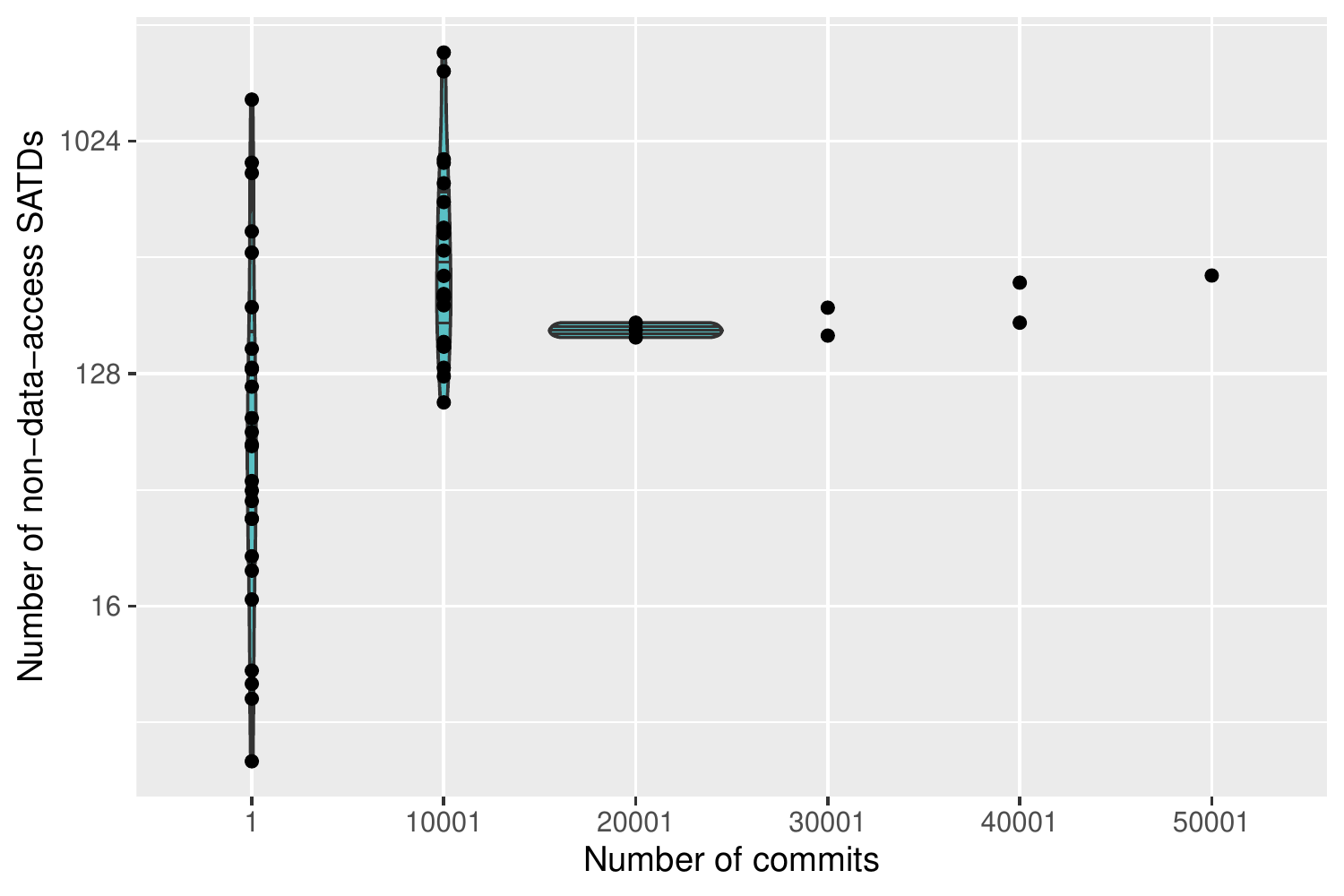}
    }\hfill
    \subfloat[Data-access SATD\label{fig:rq1dag3}]{%
        \centering
        \includegraphics[width=0.49\textwidth]{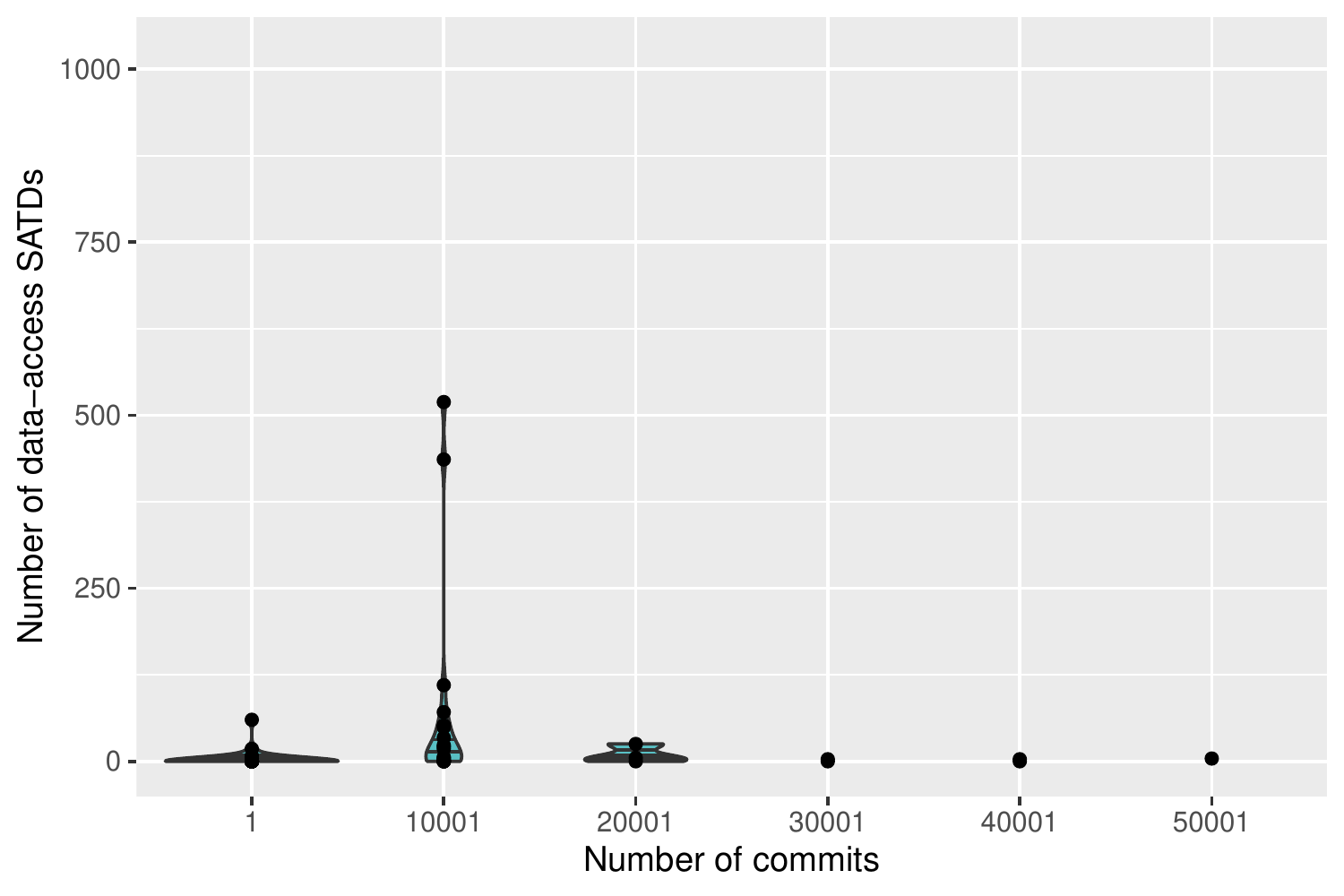}
    }
    \caption{Prevalence of regular and data-access SATD in $Group_3$.}
    \label{fig:rq1g3}
\end{figure}

\figref{fig:rq1g3} shows the distribution of regular and data-access SATD in $Group_3$.
We only have SQL systems in $Group_3$.
After commit 10,001, we have two projects where we observe SATD, and only one project, WordPress-Android,\footnote{https://github.com/wordpress-mobile/WordPress-Android} remains after commit 20,001.
The violin plot is not needed for such cases.
\figref{fig:rq1ndag3} shows that the number of regular SATDs rises between commit 1 (median=67.5) and commit 10,001 (307), then decreases at 20,001 (189).
The most significant regular SATDs (2,263) were observed at version 10,001 in \textit{ControlSystemStudio},\footnote{https://github.com/ControlSystemStudio/cs-studio} a repository of applications to operate large-scale industrial control systems.
In \figref{fig:rq1dag3}, we can see an increasing median number of data-access SATDs (0, 18) at commits 1 and 10,001.

\begin{tcolorbox}[colback=white, colframe=black,left=2pt,right=2pt,top=3pt,bottom=1pt]
\textbf{Summary:}
Data-access SATD has lower prevalence than regular SATD in both SQL and NoSQL subject systems.
We observed that the number of data-access SATDs tends to increase as systems evolve, regardless of the database type.
In most cases, NoSQL systems have higher median data-access SATD compared to SQL systems.
\end{tcolorbox}

% ----------------------------------------------------------
\subsection{RQ2: \RQTwo}
% ----------------------------------------------------------

We conducted a survival analysis to see the \newtext{persistence} of data-access SATDs. In particular, we plot the Kaplan-Meier curve for both SQL and NoSQL systems.

\begin{figure}[!ht]
\centering
    \includegraphics[width=0.7\textwidth]{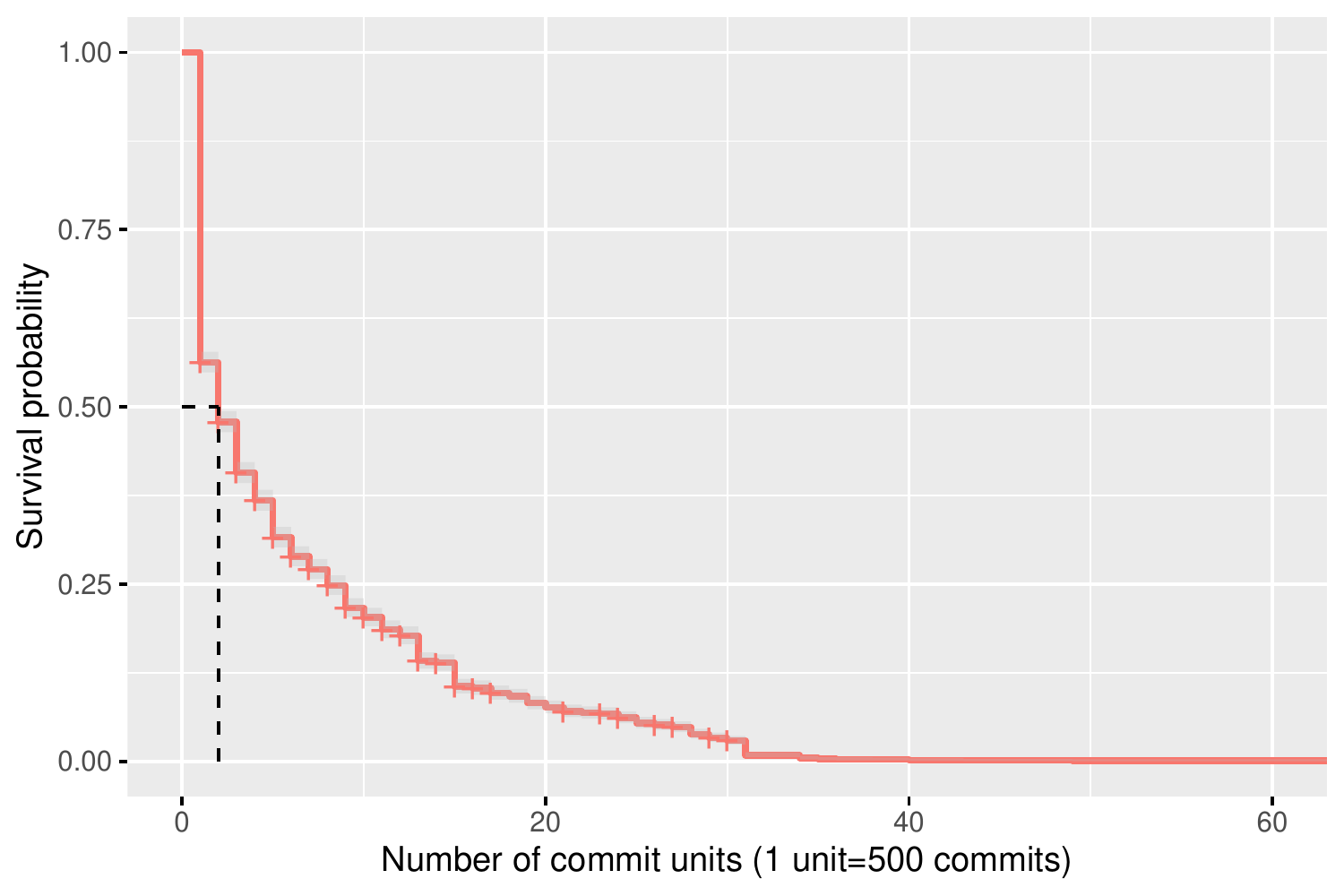}
    \caption{Kaplan–Meier survival curve for data-access SATDs in SQL subject systems. The x-axis is the number of commits. The censoring time and confidence intervals are marked on the plot. The Logrank test's p-value is indicated.}
    \label{fig:rq3sql}
\end{figure}

\figref{fig:rq3sql} shows the survival probability of data-access SATDs in SQL projects.
The median survival is 1,000 commits. \newtext{Given the average value of 500 commit time span of 535 days for SQL subject systems, described in Subsection \ref{subsec:time-metrics}, the average median survival time is 2.93 years.}
The steeper slope before 10,000 commits has two potential explanations.
One possibility is that several data-access SATDs are fixed/censored at the early stages of the projects.
Alternatively, several subject systems have a small number of commits.
The distribution of the total number of commits (median=3,729, mean=7,005, skewness=3.07) of SQL subject systems is right-skewed.
Hence, the steep slope is not likely due to small project activities.
The number of ``data-access SATD fixed'' events is 3,914, with the remaining 608 being censored.
This shows that many SATD comments are introduced and fixed at the early stages of the projects. 

\begin{figure}[!ht]
\centering
    \includegraphics[width=0.7\textwidth]{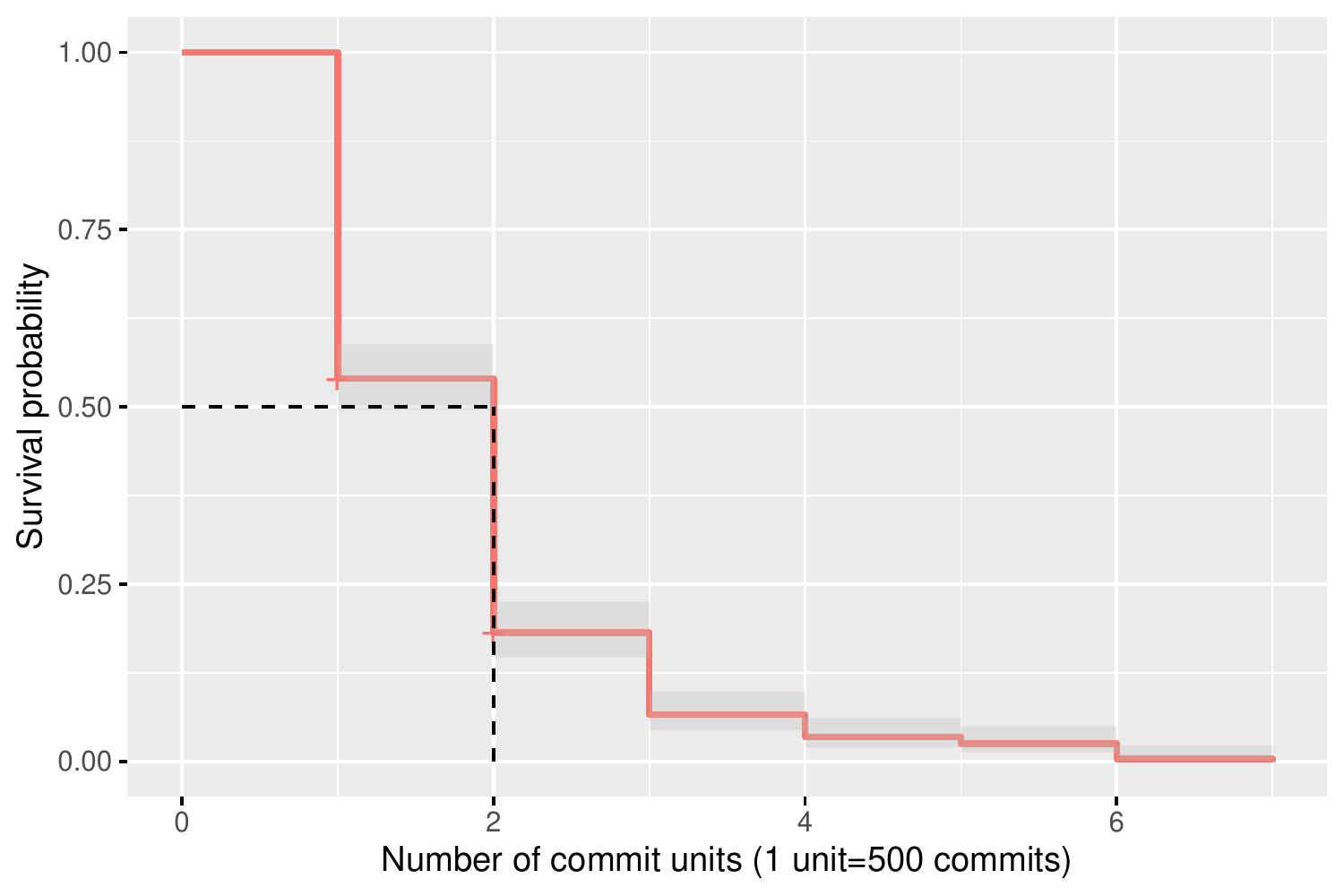}
    \caption{Kaplan–Meier survival curve for data-access SATDs in NoSQL subject systems. The x-axis represents the number of commits. The censoring time and confidence interval are marked on the plot. The Logrank test's p-value is indicated.}
    \label{fig:rq3NoSQL}
\end{figure}

\figref{fig:rq3NoSQL} shows the survival probability of data-access SATDs in NoSQL subject systems.
The median survival time of NoSQL data-access SATDs is 1,000 commits (\newtext{2.3 years using an average 500 commit time span for NoSQL subject systems as described in Subsection \ref{subsec:time-metrics}}).
The number of events is 391 out of 441, with the remaining data being censored.
The number of commits of NoSQL projects has a right-skewed distribution (median=1,927, mean=2,114, skewness=2.16).
The smaller median survival value aligns with the smaller median number of commits of NoSQL subject systems.

Many data-access SATD comments are introduced in the first versions of the systems, and several of them persisted until the latest versions.
For SQL systems, 223 (4.93\%) comments were introduced in the first version, and 152 (68.16\%) persisted until the latest version.
For NoSQL systems, 31 (7.02\%) comments were introduced in the first version, out of which 12 (38.7\%) lasted in all versions.

\begin{figure}[!ht]
\centering
    \includegraphics[width=0.7\textwidth]{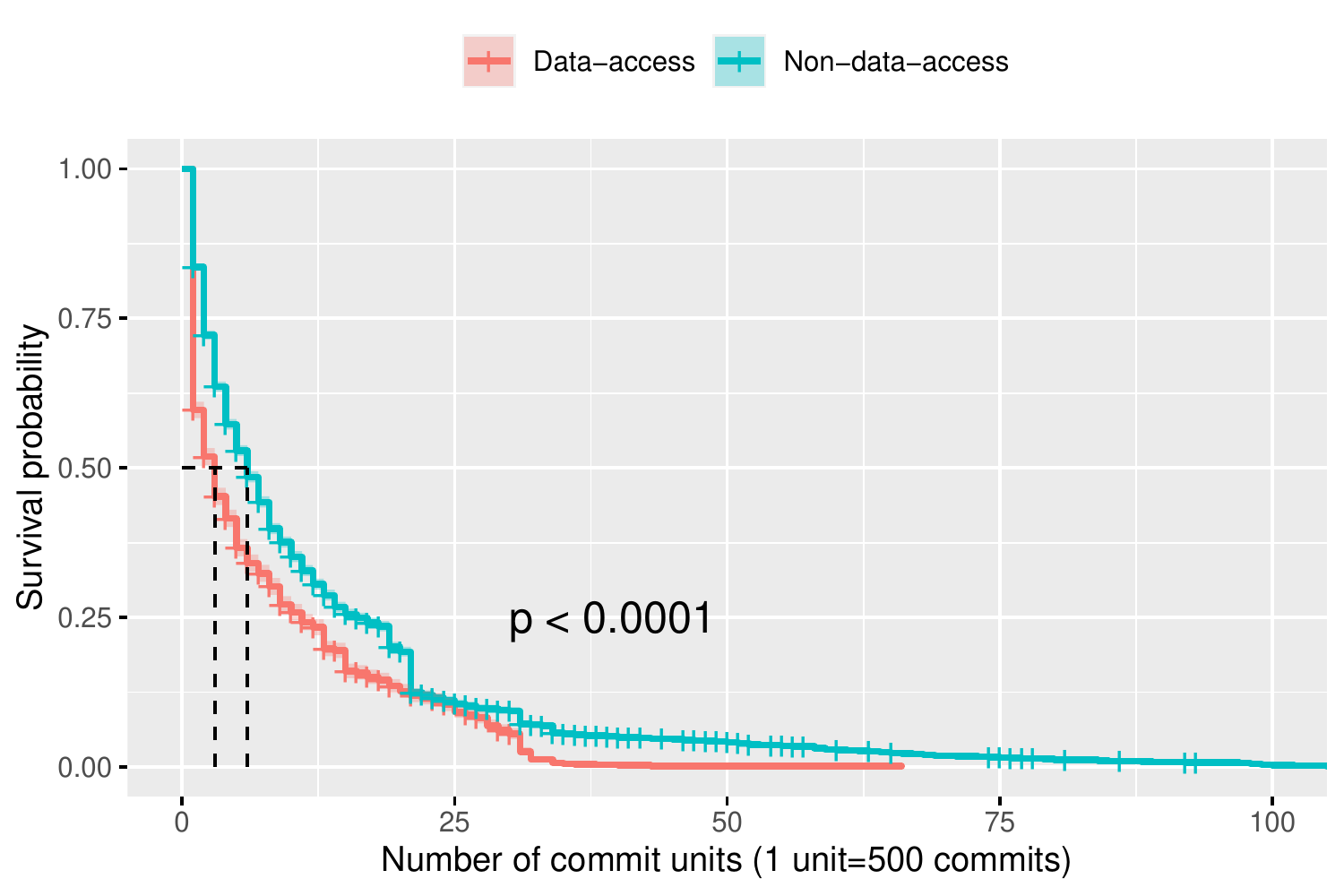}
     
    \caption{Kaplan–Meier survival curve for SQL subject systems by grouping them into data-access and regular SATD comments. The x-axis represents the number of commits. The censoring time is marked on the plot.}
    \label{fig:rq3kmnosqlcomb}
   
\end{figure}

\figref{fig:rq3kmnosqlcomb} compares the survival curves of data-access and regular SATD comments in SQL systems.
This comparison provides an insight into the prioritization of addressing technical debt. 
Data-access comments have a lower survival curve compared to their regular counterparts.
We run the Log-Rank test to compare the survival curves statistically.
The p-value of the log-rank test is $<2e-16$.
Hence, we can reject the null hypothesis that there is no difference between the survival curves of data-access and regular SATD comments.
Data-access SATDs tend to get more priority in addressing compared to regular SATDs.

\begin{figure}[!ht]
\centering
    \includegraphics[width=0.7\linewidth]{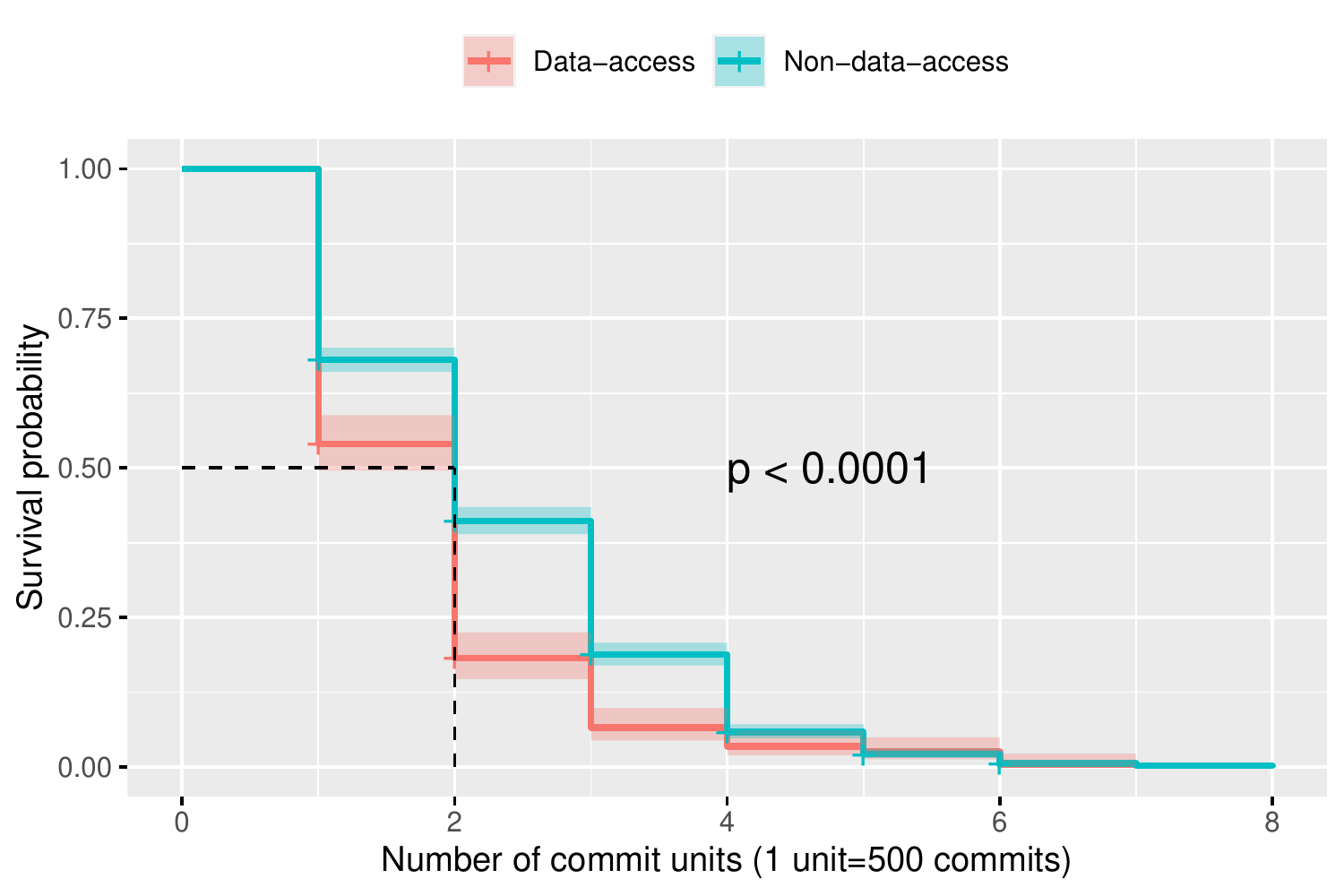}
     
    \caption{Kaplan–Meier survival curve for NoSQL subject systems by grouping them into data-access and regular SATD comments. The x-axis represents the number of commits. The censoring time is marked on the plot. }
    \label{fig:rq3kmnonsqlcomb}
\end{figure}

Similarly, \figref{fig:rq3kmnonsqlcomb} shows for NoSQL subject systems that data-access SATDs tend to get fixed quicker than regular SATDs. The Log-Rank test's p-value was $<2e-16$. Hence, we can reject the null hypothesis that there is no difference in survival curves.

\begin{tcolorbox}[colback=white, colframe=black,left=2pt,right=3pt,top=1pt,bottom=1pt]
\textbf{Summary:}

We found statistically significant differences between the survival curves of data-access and regular SATDs in both SQL and NoSQL systems, which indicates that data-access SATDs are fixed sooner than regular SATDs.
However, we also found a significant number of data-access SATDs introduced in the first versions of the systems (5\% for SQL and 7\% for NoSQL systems). Many persisted until the latest versions (68\% for SQL and 39\% for NoSQL).
\end{tcolorbox}

\subsection{RQ3: \RQThree}

In this section, we describe the result of our manual classification of SATD comments in the data-access classes.
\figref{fig:rq2Hirarchy} shows the taxonomy we extended from the work of Bavota \etal \cite{bavota2016large}.
In particular, we added a new high-level category called \textit{data-access debt} and provided more specialized categories for code debt, test debt and documentation debt.
While our primary focus is on the newly added categories, especially on the data-access debt categories, we also provide a brief description of the original categories \cite{bavota2016large} for completeness. 

\subsubsection{Distribution of manually categorized data-access SATDs}

We have manually classified 361 data-access SATD comments that represent our entire dataset with 95\% confidence.
We did not have enough information from the comments and the source code in some cases.
We labeled such comments as \textit{unclear}.
Excluding 105 \textit{false positives}, 4 multi-label and 12 unclear comments, we ended up with 240 data-access SATD comments.
\tabref{tab:rq2data} shows the distribution of the final labels in the sample dataset.
The comments under each category were presented separately for SQL and NoSQL subject systems.
We mark SATDs related to database accesses with a database icon (\faDatabase) and regular SATDs with a file icon (\faFileO).
The categories are sorted according to the \textit{total number of comments}.

\tabref{tab:rq2data} shows that a large portion of the comments belongs to sub-categories of \textit{code debt}, \textit{requirement debt} and \textit{defect debt}.
This is a similar observation with Bavota et al. \cite{bavota2016large}.
We can also see that \textit{data-access debts} are also found in smaller quantities compared to the traditional SATDs. The most considerable \textit{data access debt} is \textit{data access test debt}, followed by \textit{query construction}. 

When we contrast SATDs between SQL and NoSQL systems, we can see that most categories have a higher occurrence in SQL systems than in NoSQL systems.

\begin{landscape}
\begin{figure}[!ht]
    \includegraphics[width=\linewidth]{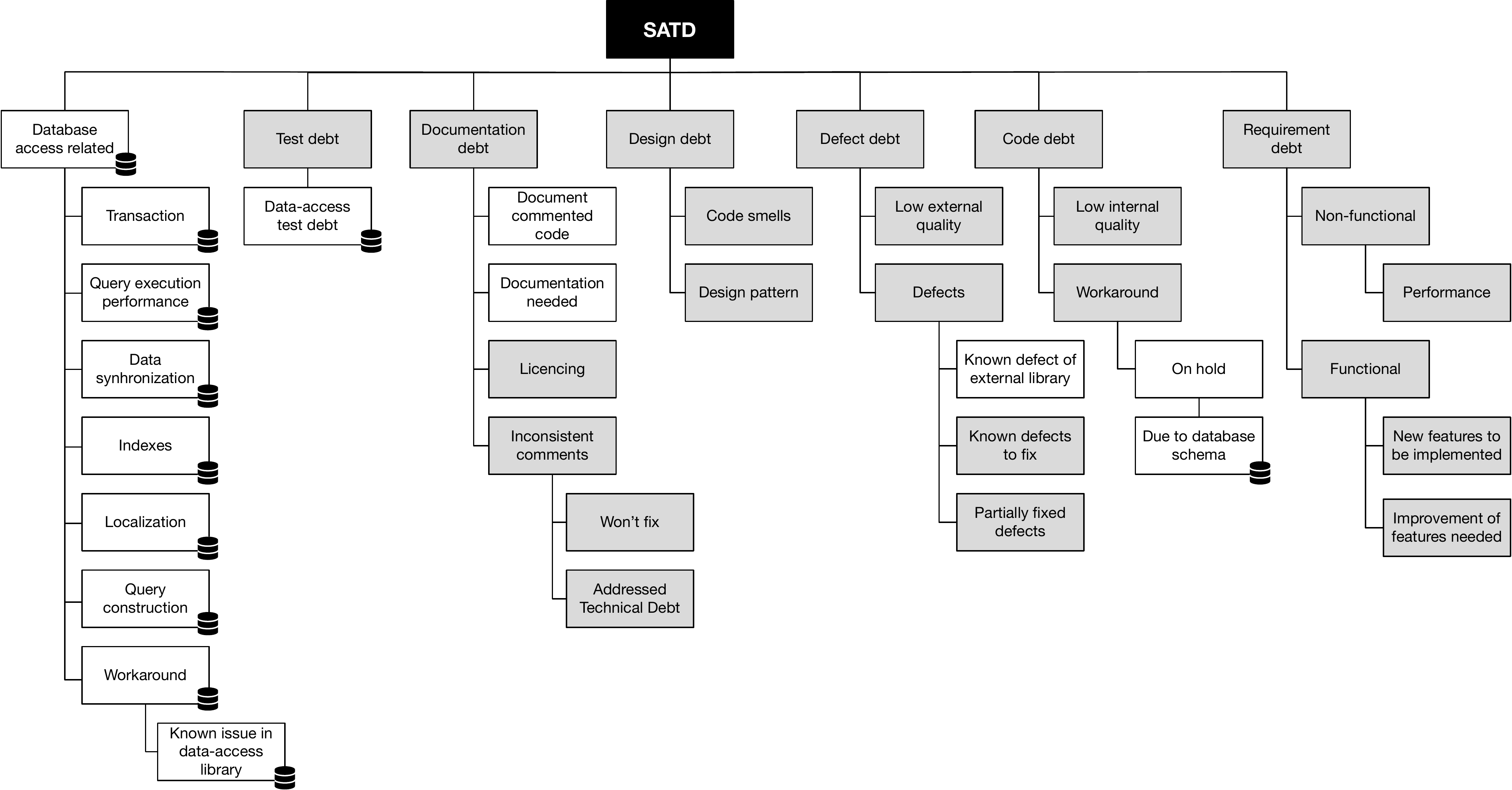}
    \caption{SATD classification hierarchy extended from Bavota \etal \cite{bavota2016large}. White boxes are newly added categories to existing categories (gray boxes). Boxes marked with a database icon (\faDatabase) are categories closely related to database accesses.} 
    \label{fig:rq2Hirarchy}
\end{figure}
\end{landscape}

Next, we describe the composition of SATDs categorized in \figref{fig:rq2Hirarchy} in the following paragraphs.
We start with the SATDs identified by Bavota \etal \cite{bavota2016large}, then we move to the newly added categories.

\begin{table}[!ht]
\caption{Distribution of categories in the manually classified dataset}
\centering
\begin{tabular}{@{}lrrrr@{}}
\toprule
\textbf{Category} & \textbf{SQL} & \textbf{NoSQL} & \textbf{Total} & \textbf{Percent} \\ \midrule
\faFileO~Low internal quality & 21 & 19 & 40 & 16.39 \\
\faFileO~Improvement to features needed & 16 & 14 & 30 & 12.30 \\
\faFileO~Known defects to fix & 9 & 16 & 25 & 10.25 \\
\faFileO~Workaround & 12 & 11 & 23 & 9.43 \\
\faFileO~New features to be implemented & 13 & 8 & 21 & 8.61 \\
\faFileO~Low external quality & 15 & 3 & 18 & 7.38 \\
\faFileO~Code smells & 10 & 6 & 16 & 6.56 \\
\faFileO~Test debt & 3 & 12 & 15 & 6.15 \\
\faDatabase~Data-access test debt & 5 & 3 & 8 & 3.28 \\
\faDatabase~Query construction & 6 & 1 & 7 & 2.87 \\
\faFileO~Document commented code & 1 & 4 & 5 & 2.05 \\
\faFileO~On hold & 1 & 4 & 5 & 2.05 \\
\faDatabase~Query execution performance & 3 & 2 & 5 & 2.05 \\
\faFileO~Performance & 1 & 2 & 3 & 1.23 \\
\faFileO~Addressed technical debt & 2 & 1 & 3 & 1.23 \\
\faFileO~Documentation needed & 3 & 0 & 3 & 1.23 \\
\faDatabase~Known issue in data access library & 1 & 1 & 2 & 0.82 \\
\faDatabase~Data synchronization & 2 & 0 & 2 & 0.82 \\
\faDatabase~Transactions & 1 & 1 & 2 & 0.82 \\
\faFileO~Known defect of external library & 1 & 1 & 2 & 0.82 \\
\faFileO~Partially fixed defects & 0 & 1 & 1 & 0.41 \\
\faDatabase~Due to database schema & 1 & 0 & 1 & 0.41 \\
\faDatabase~Localization & 1 & 0 & 1 & 0.41 \\
\faDatabase~Indexes & 0 & 1 & 1 & 0.41 \\
\faFileO~Design patterns & 1 & 0 & 1 & 0.41 \\ \bottomrule
\end{tabular}
\label{tab:rq2data}
\end{table}

\textbf{Code debt}: \textit{Code debt} includes ``\textit{problems found in the source code which can affect negatively the legibility of the code making it more difficult to be maintained}'' \cite{Alves2014}.
It is divided into \textit{low internal quality} and \textit{workaround} categories.
SATD comments that mention code quality issues related to program comprehension are categorized as \textit{low internal quality}.

For example, a comment from the \textit{low internal quality} category in \textit{Blaze-Persistence}\footnote{Blaze-Persistence, \url{https://bit.ly/3qGaXbb}} says:
\codecomment{TODO this is ugly think of a better way to do this}

Comments justified by the developers as a workaround to address specific requirements are categorized under \textit{workaround}.
For example, quick fixes that mention a hack or workaround belong to this category.
We extended \textit{workaround} SATDs with a \textit{workaround on hold} category.
An ``on-hold'' SATD comment describes a problem that can be fixed once an issue referenced in the comment is addressed \cite{maipradit2020wait}.   
 
We found a specific case of an ``on-hold'' SATD when the issue holding back the developers was due to synchronization problems with the database schema.
We dedicated the \textit{workaround on hold due to database schema} category for similar SATDs.
As an example, the comment in \textit{OpenL Tablets}\footnote{OpenL Tablets, \url{https://bit.ly/3sioFkX}} says:
\codecomment{TODO It should be removed when the table can be resolved by the ID}

\textbf{Defect debt}: 
Comments that mention bugs or defects that should be fixed but are postponed to another time are categorized under \textit{defect debt}.
The main causes of this debt can be \textit{defects} or \textit{low external quality} issues.

\textit{Defects} are further divided into \textit{known defects to fix} and \textit{partially fixed defects}.
An example of a \textit{partially fixed defect} can be seen in \textit{Snowstorm}:\footnote{Snowstorm, \url{https://bit.ly/3dEUMXN}}
\codecomment{TODO Remove this partial ESCG support}

We found two specific cases when the issue was due to a \textit{known defect of external library}; thus, we introduced a sub-category for these cases.
\textit{Low external quality} SATD comments describe problems with a high probability of becoming a bug or defect \cite{bavota2016large}, as they may affect user experience.

\textbf{Design debt}:
SATDs related to \textit{code smells} or \textit{design patterns} are grouped in this category.

Comments that discuss the violation of object-oriented design or mention refactoring as a solution are categorized under \textit{code smells}.
Comments suggesting the usage of a design pattern are classified under \textit{design patterns}. 

\textbf{Documentation debt}:
This type of SATD can be identified in comments by looking for ``\textit{missing, inadequate, or incomplete documentation of any type}'' \cite{Alves2014}.
Comments referring to issues already addressed are also categorized under documentation debt.
This might happen when developers forget to update the documentation or comments after some source code changes.
\textit{Documentation debt} is divided into \textit{inconsistent comments} and \textit{licensing} categories.
\textit{Inconsistent comments} are further divided into \textit{addressed technical debt} and \textit{won't fix} categories \cite{bavota2016large}. 

We added two new sub-categories, \textit{document commented code} and \textit{documentation needed}, as we found multiple instances of such cases. 
\textit{Document commented code} comments explain the rationale of code that was commented out but still needed due to a pending ``todo'' or ``fixme.''
Comments labeled as \textit{documentation needed} mention the necessity of providing documentation to a piece of code.

\textbf{Requirement debt}:
Comments that describe the need for new features to be implemented are categorized under \textit{requirement debt}.
Bavota \etal \cite{bavota2016large} further classified these to \textit{functional} and \textit{non-functional} requirement debt.
\textit{Functional} requirement debt includes the \textit{new feature to be implemented} and \textit{improvement to features needed} categories.

Additionally, under \textit{non-functional} requirement SATDs, we also observed a few issues related to \textit{performance} requirements.

\textbf{Test debt:}
\textit{Test debt} affects the quality of testing activities \cite{Alves2014}.
These comments are typically found in testing classes and indicate low quality of testing code, \eg in terms of readability or the appropriateness of test cases and testing conditions. 

We identified several \textit{test debt} comments in the test code related to data accesses.
We grouped these under the \textit{data access test debt} category.
Examples of these are related to the testing of database access operations such as transactions and query syntax. 
For example, a comment in \textit{Sqlg}\footnote{Sqlg, \url{https://bit.ly/3wxbAqW}} says:
\codecomment{TODO this really should execute limit on the db and finally in the step. That way less results are returned from the db}

The comment follows a query in a test method of the \textit{TestRangeLimit} class.
Sqlg provides graph computing capabilities on SQL databases, and the method tests the range specification of a query.
As the comment suggests, the query in the test could be optimized to return fewer results.

\subsubsection{Database access related SATDs}

We added \textit{database access related} as a new category that groups together SATDs dealing with the implementation of data-access logic.
This category is further divided into sub-categories.
We describe each sub-category and provide examples from the subject systems.

\textbf{Query execution performance}:
We found SATD comments dealing with issues about the execution performance of database queries.
For example, a comment in \textit{GnuCash Android}\footnote{GnuCash Android, \url{https://bit.ly/37H1PeV}} says:
\codecomment{Relies ON DELETE CASCADE takes too much time}

The comment belongs to a method that deletes all accounts and transactions from the database.
As the developers note, the cascade operation takes too much time and affects the method's performance.

\textbf{Transactions}:
We identified comments about code that deal with transactions or rollback operations.
An example of this type of debt was found in \textit{Sqlg}:\footnote{Sqlg, \url{https://bit.ly/3pRCOnK}}
\codecomment{TODO undo this in case of rollback?}

The comment appears in a method that removes a schema from a database.
The operation is performed in a transaction; however, the implementation does not undo the operation in case of a rollback.

\textbf{Workaround on known issue in data-access library}:
We found comments that described workarounds of problems existing in the data-access libraries.
In such comments, the developers explicitly reference the issue pointing to the library's issue tracking system.

The following comment in \textit{Foxtrot}\footnote{Foxtrot, \url{https://bit.ly/3urWcey}} explains a workaround by directing the developer to an issue of Hazelcast, a key-value store implementation.
\codecomment{HACK::Check https://github.com/hazelcast/hazelcast/issues/1404}

\textbf{Data synchronization}:
These SATD comments describe a synchronization issue between the application and the database.
An example comment can be found in \textit{UPortal}:\footnote{UPortal, \url{https://bit.ly/3qY50X2}}
\codecomment{todo Figure out if we should instead return the id of the system user in the DB}

The comment appears in a method called \textit{getUserIdForUsername(...)} that is supposed to return a user's ID.
However, as an additional comment says, the method ``\textit{returns 0 consistent with prior import behavior, not the id in the database.}''

\textbf{Indexes}:
Comments about issues related to indexes in the database are grouped under this category.
For example, the following comment in \textit{Sqlg}\footnote{Sqlg, \url{https://bit.ly/3aIqEcc}} describes the need for support for indexes. 
\codecomment{TODO Sqlg needs to get more sophisticated support for indexes i.e. function indexes on a property etc.}

\textbf{Localization}:
We found comments about localization issues in the database, \ie problems with character sets or collation.
The following comment in \textit{Robolectric}\footnote{Robolectric, \url{https://bit.ly/3umvXpD}} highlights the need for creating a collator as part of registering a localized collator. 
\codecomment{TODO: find a way to create a collator^^J%
  // http://www.sqlite.org/c3ref/create_collation.html^^J%
  // xerial jdbc driver does not have a Java method for sqlite3_create_collation}

\textbf{Query construction}:
We found comments that mentioned issues about the construction of database queries.
The following comment in \textit{Carbon-apimgt}\footnote{Carbon-apimgt, \url{https://bit.ly/2NvDZvQ}} notes a pending task to filter results by the status of the APIs.
\codecomment{TODO FILTER RESULTS ONLY FOR ACTIVE APIs}

The query marked with the todo comment returns unnecessary records when only a specific API context is needed.

\begin{tcolorbox}[colback=white, colframe=black,left=2pt,right=3pt,top=1pt,bottom=1pt]
\textbf{Summary:}
We identified in data-access classes a large variety of SATD categories from the taxonomy of Bavota \etal \cite{bavota2016large}.
Low internal quality code debt has the highest prevalence among data-access SATDs in both SQL and NoSQL subject systems.
Most of the data-access SATDs have a higher prevalence in SQL subject systems compared to NoSQL subject systems.
Besides the categories in the taxonomy of Bavota \etal \cite{bavota2016large}, we found several SATDs pertaining to data access operations such as query construction, data synchronization, index management and transactions.
\end{tcolorbox}

\subsection{RQ4: \RQFour}
\label{sec:design_rq4}

In this subsection, we present our result and analysis concerning the circumstances behind the introduction and removal of data-access SATDs.
We first discuss when data-access SATDs are introduced and removed.
Then we discuss the reasons motivating their introduction and removal.

\subsubsection{When are data-access SATDs introduced?}

To answer this question, we analyze the change history of the data-access SATDs labeled in RQ3 and identify the commits introducing the comments.
The commit at which the comment first appears in change history is referred to as the \textit{data-access SATD introducing commit}.

We measure the \textit{introduction time} as the number of commits it takes for SATDs to manifest in the subject systems.
As for the survival analysis, we use the number of commits to measure time as it is more reflective of software development activities.
Similarly, we define \textit{removal time} as the number of commits between the SATD comment’s introduction and removal.
Since each system's number of commits varies, we normalized the introduction time and removal time using \eqref{eq:introtime} (see \secref{sec:design_rq4}).

\begin{figure}[!ht]
\centering
    \includegraphics[width=0.5\textwidth]{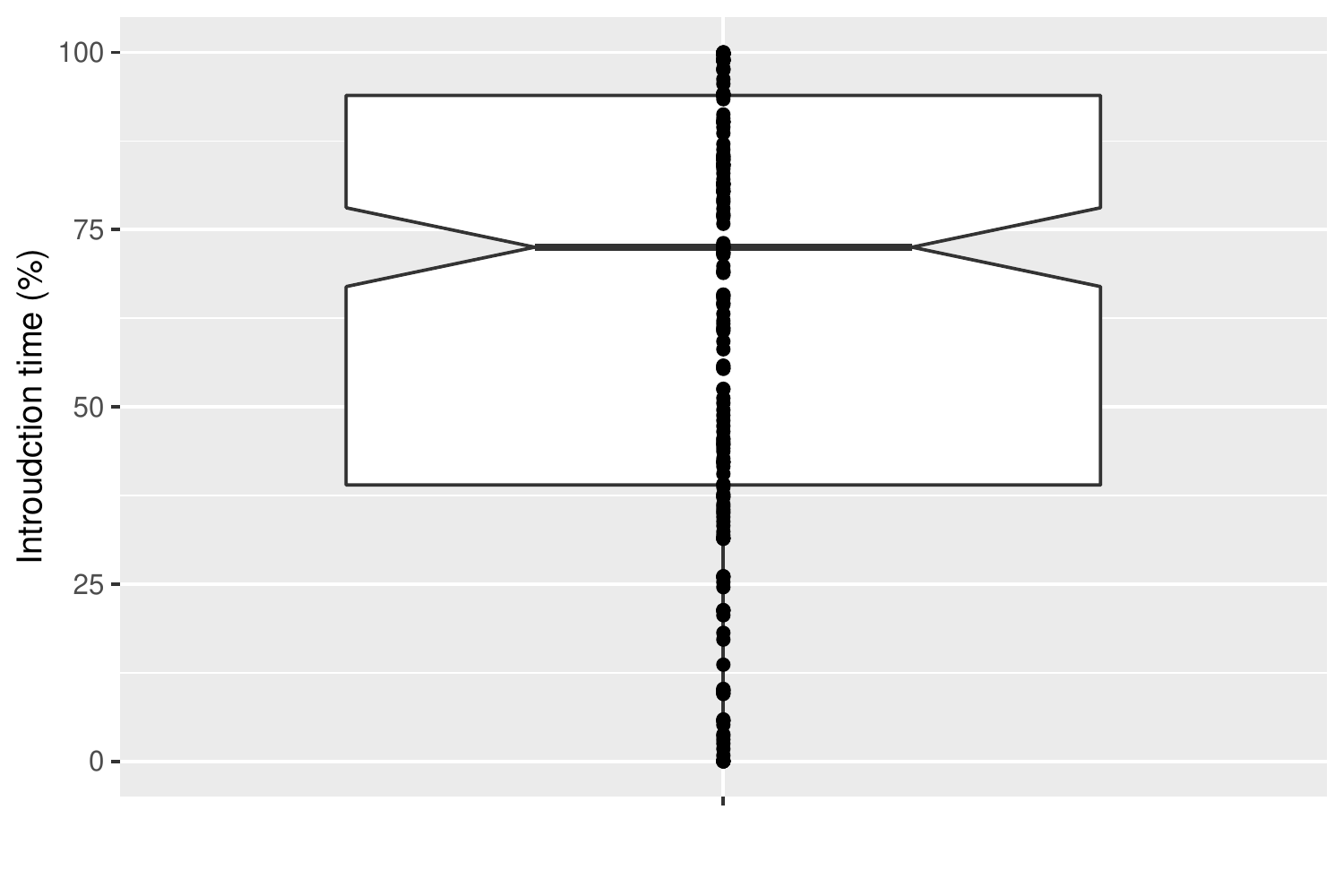}
    \caption{Distribution of data-access SATD introduction time}
    \label{fig:rq4tintroAll}
\end{figure}

\figref{fig:rq4tintroAll} shows the overall distribution of data-access SATD introduction time.
The distribution is right-skewed with the median introduction time (72.53\%) and mean (64.14\%).
This indicates that most of the data-access SATD introducing commits did not happen at the beginning of the change history.
This also confirms our observation of the survival analysis in RQ2.
Data-access SATDs seem to be introduced at later stages in the change history.
We also identified SATDs committed in the most recent snapshots of the subject systems (introduction time=100\%).

\begin{figure}[!ht]
\centering
        \includegraphics[width=0.5\textwidth]{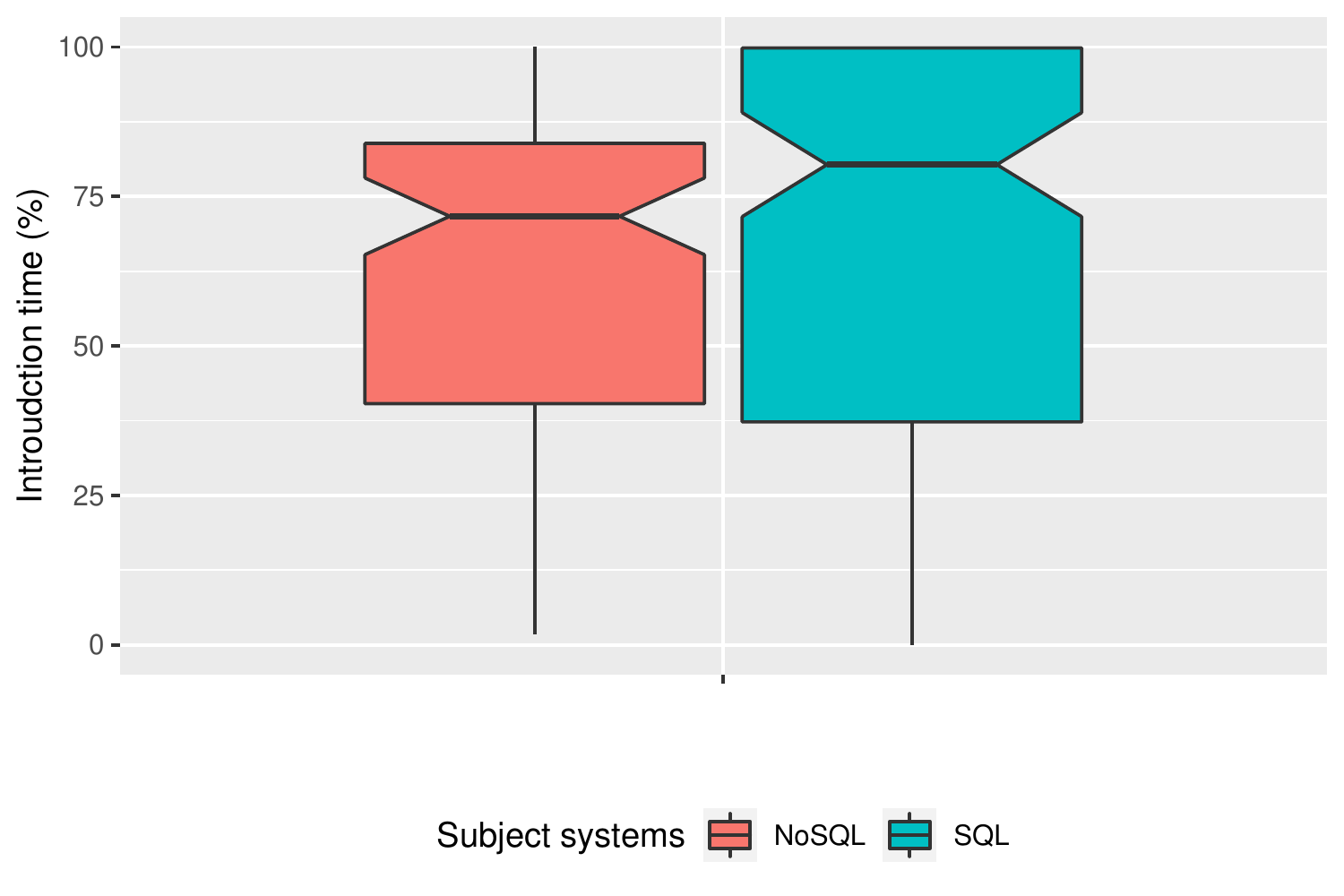}
     
    \caption{Distribution of data-access SATD introduction time in SQL and NoSQL subject systems}
    \label{fig:rq4tsqlvsnosql}
\end{figure}

\figref{fig:rq4tsqlvsnosql} shows the distribution of introduction time for SQL and NoSQL systems.
For both SQL and NoSQL data-access SATDs, introduction time is right-skewed.
The notches of the SQL and NoSQL overlap, which means that the difference in the median is not significant.
SQL data-access SATDs have a slightly higher median (80.31\%) than in NoSQL systems (71.67\%).

\tabref{tbl:rq4tcats} shows the number of comments, mean, and median introduction time for all data-access SATD categories.
The categories are ordered by the median introduction time from highest to lowest.
\textit{Low external quality} and \textit{design patterns} data-access SATDs are introduced in the latest stages of change history among all the categories.
On the other extreme, most of the database access related SATDs tend to be introduced at the early stages of change history.
Compared to regular SATDs, most of the database access related SATDs are introduced earlier.
\textit{Transactions}, \textit{indexes}, and \textit{data-access test debt} tend to be introduced at later stages.
\textit{Addressed technical debt} comments tend to be introduced at the very beginning of the subject systems' development.  

\begin{table}[!ht]
\caption{Data-access SATD introduction time for SATD categories}
\centering
\begin{tabular}{@{}lrrr@{}}
\toprule
\textbf{Category} & \textbf{Comments} & \textbf{Mean} & \textbf{Median} \\ \midrule
\faFileO~Low external quality & 18 & 81.34& 99.83 \\
\faFileO~Design patterns & 1 & 99.83 & 99.83\\
\faFileO~Performance & 3 & 87.86 & 82.96 \\
\faFileO~Workaround & 23 & 62.76 & 82.12\\
\faDatabase~Data-access test debt & 8 & 74.76 & 81.31\\
\faFileO~Known defects to fix & 25 & 75.24 & 80.43 \\
\faFileO~New features to be implemented & 21 & 64.25 & 80.43 \\
\faFileO~Code smells & 16 & 72.08 & 77.28 \\
\faDatabase~Transactions & 2 & 72.62 & 72.62 \\
\faDatabase~Indexes & 1 & 72.53 & 72.53 \\
\faFileO~Document commented code & 5 & 49.15 & 72.31 \\
\faFileO~Test debt & 15 & 71.10 & 71.67 \\
\faFileO~Improvement to features needed & 30 & 59.89 & 70.66 \\
\faFileO~Known defect of external library & 2 & 68.15 & 68.15 \\
\faFileO~Multi-label & 4 & 63.09 & 64.26 \\
\faDatabase~Localization & 1 & 61.19 & 61.19 \\
\faFileO~Low internal quality & 40 & 56.84 & 58.75 \\
\faFileO~Documentation needed & 3 & 66.39 & 50.54 \\
\faFileO~On hold & 5 & 46.73 & 48.09\\
\faDatabase~Due to database schema & 1 & 47.30 & 47.30 \\
\faDatabase~Data synchronization & 2 & 45.64 & 45.64 \\
\faDatabase~Query execution performance & 5 & 48.93 & 44.68 \\
\faDatabase~Known issue in data access library & 2 & 43.57 & 43.57 \\
\faDatabase~Query construction & 7 & 48.08 & 37.29 \\
\faFileO~Partially fixed defects & 1 & 21.30 & 21.30 \\
\faFileO~Addressed technical debt & 3 & 28.90 & 2.51 \\ \bottomrule
\end{tabular}
\label{tbl:rq4tcats}
\end{table}

\subsubsection{When are data-access SATDs removed?}

\begin{figure}[!ht]
\centering
        \includegraphics[width=0.5\textwidth]{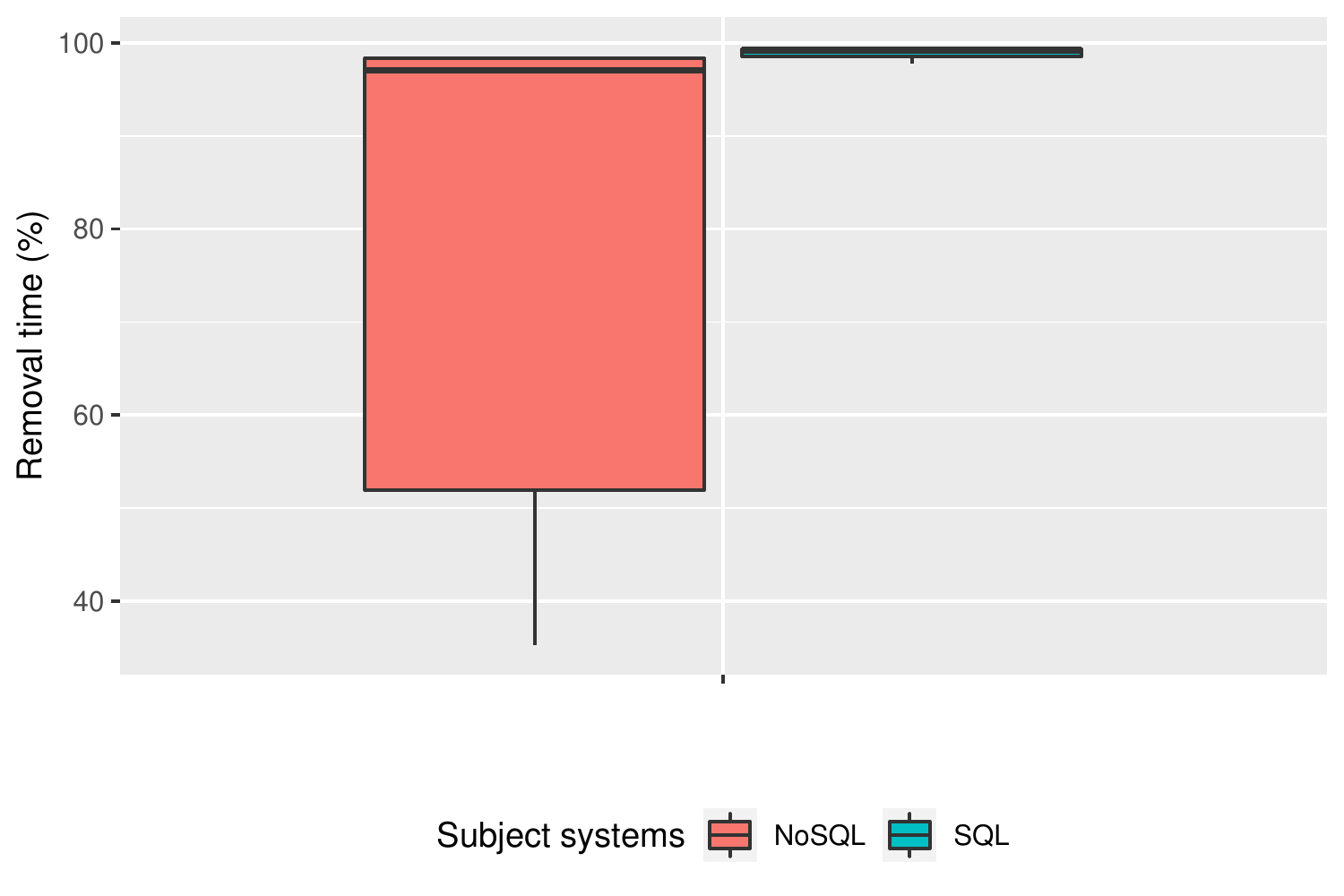}
     
    \caption{Distribution of data-access SATD removal time in SQL and NoSQL subject systems}
    \label{fig:rq4trsqlvsnosql}
\end{figure}

We found 12 data-access SATDs that were removed at different stages of the change history.
\figref{fig:rq4trsqlvsnosql} shows the distribution of data-access SATD removal time for SQL and NoSQL subject systems.
Both SQL and NoSQL SATDs were removed at the latter stages close to the most recent versions.
The median removal time is 99.58\% for SQL and 98.58\% for NoSQL data-access SATDs.

\begin{table}[!ht]
\caption{Distribution of data-access SATD removal time among the data-access categories}
\centering
\begin{tabular}{@{}lrrrrr@{}}
\toprule
\textbf{Category} & \textbf{Comments} & \textbf{Mean} & \textbf{Median} & \textbf{Minimum} & \textbf{Maximum} \\ \midrule
\faFileO~Improvement to features needed & 2 & 99.47 & 99.47 & 99.36 & 99.58 \\
\faFileO~Code smells & 1 & 98.54 & 98.54 & 98.54 & 98.54 \\
\faFileO~Known defects to fix & 2 & 98.19 & 98.19 & 97.80 & 98.58 \\
\faFileO~Test debt & 1 & 98.15 & 98.15 & 98.15 & 98.15 \\
\faFileO~Low internal quality & 5 & 77.34 & 97.06 & 35.29 & 99.22 \\
\faFileO~Document commented code & 1 & 47.32 & 47.32 & 47.32 & 47.32 \\ \bottomrule
\end{tabular}
\label{tbl:rq4trcat}
\end{table}

\tabref{tbl:rq4trcat} shows the distribution of data-access SATD removal time grouped by categories.
We did not have any removed comments from the \textit{database access related} SATD category.
\textit{Improvement of features needed} comments tend to be introduced at later stages of change history with the highest median removal time of 99.58\%.
On the other hand, \textit{document commented code} comments were introduced in the middle stages of the change history (median=47.32\%).

\subsubsection{Why are data-access SATDs introduced and removed?}

We now focus on the potential reasons for data-access SATDs' introduction and removal.
We manually labeled the data-access SATDs' introducing/removing commit messages to classify their purposes.
We classified the goal of the commit messages as \textit{bug fixing}, \textit{enhancement}, \textit{new feature}, \textit{refactoring}, and \textit{merging}.
Some commit messages described \textit{multiple goals}, and some comments were labeled \textit{unclear} as they did not contain enough information in the commit message for categorization. 

\begin{table}[!ht]
\caption{Data-access introducing commit goals in NoSQL and SQL subject systems}
\centering
\resizebox{\textwidth}{!}{%
\begin{tabular}{@{}llllllll@{}}
\toprule
\textbf{Systems} & \textbf{Bug Fixing} & \textbf{Enhancement} & \textbf{Multiple Goals} & \textbf{New Feature} & \textbf{Refactoring} & \textbf{Unclear} & \textbf{Merging} \\ \midrule
NoSQL & 17 & 22 & 3 & 30 & 38 & 5 & 0 \\
SQL & 45 & 6 & 3 & 32 & 38 & 1 & 4 \\ \bottomrule
\end{tabular}%
}
\label{tbl:rq4wall}
\end{table}

\tabref{tbl:rq4wall} summarizes the various goals of data-access SATDs' introductions.
Considering NoSQL data-access SATDs, \textit{refactoring} is the most associated reason with 38 instances (33.04\%).
It is followed by \textit{new feature} with 30 cases (26.09\%) and 22 \textit{enhancements} (19.13\%).
For SQL, \textit{bug fixing} was the most often mentioned reason in comments with 45 instances (34.88\%).
It is followed by \textit{refactoring} with 38 cases (29.46\%) and 30 \textit{new features} (23.26\%).
Overall, \textit{bug fixing} and \textit{refactoring} are the main reasons behind the introduction of data-access SATDs. 

\begin{table}[!ht]
\caption{Data-access SATD introducing commit goals grouped by data-access SATD categories}
\centering
\resizebox{\textwidth}{!}{%
\begin{tabular}{@{}lrrrrrrr@{}}
\toprule
\textbf{Categories} & \textbf{Bug Fixing} & \textbf{Enhancement} & \textbf{Multiple Goals} & \textbf{New Feature} & \textbf{Refactoring} & \textbf{Unclear} & \textbf{Merging} \\ \midrule
\faFileO~Low internal quality & 11 & 7 & 2 & 4 & 14 & 2 & \deemph{0}\\
\faFileO~Workaround & 4 & 3 & 1 & 6 & 9 & \deemph{0}& \deemph{0}\\
\faFileO~On hold & 1 & 1 & 3 & \deemph{0}& \deemph{0}& \deemph{0}& \deemph{0}\\
\faDatabase~Due to database schema & \deemph{0}& 1 & \deemph{0}& \deemph{0}& \deemph{0}& \deemph{0}& \deemph{0}\\
\faDatabase~Query execution performance & 1 & \deemph{0}& 1 & 1 & 2 & \deemph{0}& \deemph{0}\\
\faDatabase~Transactions & 1 & \deemph{0}& \deemph{0}& 1 & \deemph{0}& \deemph{0}& \deemph{0}\\
\faDatabase~Known issue in data-access library & \deemph{0}& \deemph{0}& \deemph{0}& 1 & 1 & \deemph{0}& \deemph{0}\\
\faDatabase~Data synchronization & \deemph{0}& \deemph{0}& \deemph{0}& \deemph{0}& 2 & \deemph{0}& \deemph{0}\\
\faDatabase~Indexes & \deemph{0}& \deemph{0}& \deemph{0}& 1 & \deemph{0}& \deemph{0}& \deemph{0}\\
\faDatabase~Localization & 1 & \deemph{0}& \deemph{0}& \deemph{0}& \deemph{0}& \deemph{0}& \deemph{0}\\
\faDatabase~Query construction & \deemph{0}& 1 & \deemph{0}& 5 & 1 & \deemph{0}& \deemph{0}\\
\faFileO~Known defect of external library & 1 & 1 & \deemph{0}& \deemph{0}& \deemph{0}& \deemph{0}& \deemph{0}\\
\faFileO~Known defects to fix & 7 & 3 & 1 & 5 & 8 & 1 & \deemph{0}\\
\faFileO~Low external quality & 9 & \deemph{0}& \deemph{0}& 3 & 4 & \deemph{0}& 2 \\
\faFileO~Partially fixed defects & \deemph{0}& \deemph{0}& \deemph{0}& \deemph{0}& 1 & \deemph{0}& \deemph{0}\\
\faFileO~Code smells & 5 & 1 & \deemph{0}& 5 & 5 & \deemph{0}& \deemph{0}\\
\faFileO~Design patterns & 1 & \deemph{0}& \deemph{0}& \deemph{0}& \deemph{0}& \deemph{0}& \deemph{0}\\
\faFileO~Document commented code & 3 & \deemph{0}& \deemph{0}& 1 & \deemph{0}& 1 & \deemph{0}\\
\faFileO~Documentation needed & 2 & \deemph{0}& \deemph{0}& \deemph{0}& 1 & \deemph{0}& \deemph{0}\\
\faFileO~Addressed technical debt & 1 & \deemph{0}& \deemph{0}& \deemph{0}& 2 & \deemph{0}& \deemph{0}\\
\faFileO~Multi-label & 1 & 1 & \deemph{0}& 1 & \deemph{0}& \deemph{0}& \deemph{0}\\
\faFileO~Improvement to features needed & 7 & 2 & \deemph{0}& 10 & 10 & 1 & \deemph{0}\\
\faFileO~New features to be implemented & 2 & 3 & \deemph{0}& 5 & 9 & 1 & 1 \\
\faFileO~Performance & 1 & \deemph{0}& \deemph{0}& \deemph{0}& 2 & \deemph{0}& \deemph{0}\\
\faFileO~Test debt & 1 & 5 & 1 & 6 & 2 & \deemph{0}& \deemph{0}\\
\faDatabase~Data access test debt & 2 & \deemph{0}& 1 & 3 & 2 & \deemph{0}& \deemph{0}\\ \bottomrule
\end{tabular}%
}
\label{tab:rq4wicat}
\end{table}

\tabref{tab:rq4wicat} shows the introduction goals grouped by data-access debt categories.

In general, \textit{refactoring}, \textit{new feature} and \textit{bug fixing} appear to be the most common reasons. 
However, only considering the database access related SATDs, they are mainly introduced during \textit{refactoring}.
Another interesting observation is that \textit{code smells} are introduced during \textit{refactoring} (31.25\%), \textit{bug fixing} (31.25\%) and \textit{new feature} (31.25\%).
This means that refactoring, which is supposed to fix code smells, could also introduce other code smells and SATDs.

\begin{table}[!ht]
\caption{Data-access SATD removing commit goals grouped by data-access SATD categories}
\centering
\begin{tabular}{@{}lrr@{}}
\toprule
\textbf{Category} & \textbf{Commit Goal} & \textbf{Comments} \\ \midrule
\multirow{2}{*}{Low internal quality} & Enhancement & 3 \\
 & New Feature & 2 \\ \midrule
\multirow{2}{*}{Known defects to fix} & Enhancement & 1 \\
 & New Feature & 1 \\ \midrule
Code smells & Unclear & 1 \\ \midrule
Document commented code & Bug fixing & 1 \\ \midrule
\multirow{2}{*}{Improvement to features needed} & Bug fixing & 1 \\ 
 & New Feature & 1 \\ \midrule
Test debt & Bug fixing & 1 \\ \bottomrule
\end{tabular}%
\label{tbl:rq4wrcat}
\end{table}

We present the removal goals of SATD categories in Table \ref{tbl:rq4wrcat}.
\textit{Low internal quality} is associated with \textit{enhancement} (60\%) and \textit{new feature} (40\%).
The remaining SATD categories have 6 instances combined.

\begin{table}[!ht]
\caption{Data-access SATD removing commit goals for SQL and NoSQL subject systems}
\centering
\begin{tabular}{@{}lrrrr@{}}
\toprule
\textbf{Commit Goal} & \textbf{Enhancement} & \textbf{New Feature} & \textbf{Bug Fixing} & \textbf{Unclear} \\ \midrule
SQL & 1 & 2 & 1 & 1 \\
NoSQL & 3 & 2 & 2 & \deemph{0}\\ \midrule
\textbf{Total} & \textbf{4} & \textbf{4} & \textbf{3} & \textbf{1} \\
\bottomrule
\end{tabular}%
\label{tab:rq4wrsqlvsnosql}
\end{table}

\tabref{tab:rq4wrsqlvsnosql} summarizes the goals of the removals of data-access SATDs.
Several comments were removed for feature \textit{enhancements} and \textit{new features}.
\textit{Bug fixing} commits also contribute to the reduction of data-access SATD.
Both SQL and NoSQL systems follow a similar distribution of commit goals.

\begin{tcolorbox}[colback=white, colframe=black,left=2pt,right=3pt,top=1pt,bottom=1pt]
\textbf{Summary:}
Most SATD comments in data-access classes are introduced at the later stages of change history.
However, SATD comments where database access is explicitly mentioned (\ie database access related categories in the taxonomy) are introduced earlier than SATD comments unrelated to database accesses.
We observed similar distribution between SQL and NoSQL data-access SATDs in introduction time.
\textit{Bug fixing} and \textit{refactoring} are the main reasons behind the introduction of data-access SATDs, followed by \textit{feature enhancement} and \textit{supporting new features}.
Data-access debt removal commits are often associated with \textit{feature enhancements}, \textit{new features}, and \textit{bug fixing}.
None of the observed removal events was associated with \textit{refactoring}.
We did not find removed \textit{database access related SATD} comments. 
\end{tcolorbox}

\section{Discussion}

\label{sec:dis}
\newtext{
The goal of this study is to explore SATDs in data-intensive systems. In particular, we investigated the prevalence, persistence, composition of SATDs, and introduction and removal circumstances. The results show that SATDs are prevalent in data-intensive systems, and their prevalence increases as systems evolve. This pattern is similar to traditional software systems. Bavota and Russo~\cite{bavota2016large} showed that SATDs are prevalent and increase as new ones are introduced during software evolution. This indicates that in both traditional and data-intensive systems, developers tend to introduce new SATDs instead of addressing existing ones. In addition, our results show that the prevalence of SATDs is different between SQL and NoSQL data-intensive systems. Given that NoSQL persistence systems are getting higher preference due to the advantages they offer in terms of schema flexibility and scalability and our result showing more prevalent SATDs in some NoSQL-based systems, our findings motivate further investigation of the impact of the persistence technologies on SATD.}

\newtext{
Our results regarding the persistence of SATDs in data-intensive systems are similar to traditional systems. Bavota and Russo~ \cite{bavota2016large} found that the median survival time of SATDs to be 1000 commits for traditional software systems. We also find similar median survival times for both SQL and NoSQL subject systems. On the other hand, Maldonado \etal \cite{maldonado2017empirical} reported that SATDs persist up to 173 days on average using five open-source traditional software systems. This implies that SATDs in data-intensive systems have even higher persistence (more than two years on average in our case). We also found that a significant number of SATDs persisted in all versions without getting addressed. Since the longer the SATD stays in the system, the higher the cost of repaying, practitioners should incorporate fixing technical debts as part of their workflow. This result highlights the importance of research work in SATDs in terms of providing tool support, raising awareness of the costs of technical debts, and providing processes and frameworks for monitoring technical debt.}

\newtext{The state-of-the-art SATD detection systems do not differentiate between different types of SATDs. 
One reason for this could be the lack of information on the specific types of SATDs. In this direction, Bavota and Russo~\cite{bavota2016large} provided a taxonomy of SATDs, including design debt, code debt, defect debt, requirement debt, and test debt. While they addressed most of the software development workflow, they did not cover data-access debts since the subject systems were not data-intensive. We extended their taxonomy, incorporating 11 new database access-specific SATDs generalizing their taxonomy to data-intensive systems. This taxonomy can be utilized for proposing SATD detection approaches that provide more information than their mere existence. This, in turn, helps practitioners in their effort to manage technical debts and future researchers to investigate the impacts of specific types of SATDs on software quality. We find that low internal quality code debts were the most prevalent SATDs among our subject systems. Code debts are also found to be dominant SATDs in traditional software systems \cite{bavota2016large}. Hence, future research efforts should focus more on code debts as they are more prevalent SATDs in software systems. Data-access SATDs are also important in the context of data-intensive systems.}

\newtext{Our fine-grained analysis on data-access SATDs showed that most data-access SATD comments are introduced as the subject systems evolve rather than at the initial stages indicating that they are introduced as a result of software evolution. A software system can evolve for various reasons such as bug fixing, adding new features, improving features, and refactoring activities. Developers should take care to assess the cost of the SATD they introduce with such activities. Our results also show that the introduction of data-access SATDs is mainly associated with refactoring. However, this motivates further investigation on how and why refactoring operations are associated with SATDs. This could be done by extracting refactoring information using refactoring detection tools and co-relating with the SATD's introduced. This, in turn, leads to the development of refactoring tools that also suggest developers when to admit technical debts.
}

\section{Threats to validity}
\label{sec:ttv}
\textbf{\textbf{Threats to construct validity:}} Threats to construct validity concern the relation between theory and observation. We relied on a list of keywords and import statements to select subject systems and distinguish data-access classes (DAC) from non-data-access classes (NDC) within those systems. We may have missed some keywords and import statements, which would lead us to overestimate the set of NDCs. Conversely, it is possible that some classes are considered as DACs (\ie that import database-related packages belong to our list), but do not (directly) query the database. Hence, we may also slightly overestimate the actual set of DACs in the software systems considered. 
We checked 100 randomly selected data-access classes and found that 82\% of those directly query the database. 

Another threat to construct validity is the precision of the SATD detector tool. The 73.7\% $F_1$ score shows that the tool could introduce a significant number of false positives. Indeed, we conducted a manual analysis and identified a considerable number of false positives. However, The SATD detector is a state-of-the-art tool whose base approach was also used in other studies (\eg \cite{bavota2016large}). Improving the accuracy of the SATD detector is out of the scope of the paper. However, the conclusions from this paper are carefully formulated and need to be interpreted taking into account the imprecise nature of the tool.

There might be cases when SATD comments are removed without code changes in effect. This may mean that the SATD admitted earlier is no longer viewed as technical debt by the developers, or they may not be interested in keeping track of that SATD \cite{Zampetti2018}. Such cases are not actual removals of SATDs. Zampetti \etal \cite{Zampetti2018} conducted an empirical study on Java open-source systems and observed that such cases are not frequent ($<$ 10\%) in most cases and the maximum being 17\%. 

\newtext{We used the number of commits as a metric to measure developer activity instead of time due to the variations in commit time span across subject systems and in between different snapshots of a subject system. However, the number of commits may not accurately represent the time spent by developers on technical debt. To help mitigate this threat, we provided the typical 500 commit time span for each subject system in the replication package as an indication of time.}

\textbf{Threats to internal validity}: Internal validity concerns how one can be confident on claimed cause and effect relation. We did not claim any causation in our study. We only analyzed the diffusion and survival of SATD in SQL and NoSQL subject systems. Hence, our study is not subjected to threats to internal validity.

\textbf{Threats to conclusion validity}: Conclusion validity concerns the degree to which the statistical conclusions about the claimed relationships are reasonable. To avoid conclusion threats to validity, we only used non-parametric statistical tests. 

\textbf{Threats to external validity}: External validity concerns the generalizability of findings outside the study context. Our study considers different types of projects in terms of database technology (SQL or NoSQL), application domain, size, and the number of database interactions. We also covered projects that use different drivers and frameworks to interact with the database. We only considered Java projects for analysis. However, our investigation approach is generalizable to other programming languages.

\textbf{Threats to reliability validity}: Reliability validity concerns factors that cause an error in data collection and analysis. To minimize potential threats to reliability, we analyzed open-source projects available on GitHub and provided a replication package that contains our dataset and analysis scripts \cite{replication}.

%-----------------------------------------------------------------------
\section{Conclusion and future work}
\label{sec:conc}

Technical debt represents the costs associated with favoring short-term low-quality solutions rather than appropriate solutions that take more time.
Developers use SATD comments to track technical debt and reserve it for future fixes.
We conducted a large-scale empirical study on data-access SATD using 102 open-source data-intensive systems.
In particular, we extracted SATD comments from multiple snapshots of each subject system and \newtext{explored the prevalence and persistence of data-access SATD.}
We conducted a manual analysis on representative data-access SATDs to categorize them under the type of SATD refereed.
We further analyzed the data-access SATDs to understand the circumstances behind introducing/removing such debt.

Results show that data-access SATDs are introduced as software gets more mature, and many instances of SATDs persisted for a more extended time.
Bug fixing and refactoring are the main reasons behind the introduction of data-access SATDs followed by feature enhancements and new features.
The observed SATD removal activities are not associated with refactoring, which implies that the removals are merely parts of bug fixing or feature enhancement activities.

SATDs in general and data-access SATDs, in particular, are critical to data-intensive systems as they determine the quality of the subject systems in terms of robustness and efficiency of data-access operations.
Supporting more functionalities and maintaining code quality at the same time is a general problem to any software system.
Having the right balance would help maintain software quality and reduce technical debt costs in the long run.

Our exploratory study could be extended in different ways.
One extension could be to validate the newly identified database access-related SATDs and evaluate how developers prioritize such SATDs.
Another extension of this study could be to investigate the impact of data-access SATDs on software quality.
This could help demonstrate the harmfulness of technical debt to practitioners, which is particularly important in the context of data-intensive systems. 

\section{Declarations}
\subsection{Funding and Conflicts of interests}
The authors did not receive support from any organization for the submitted work.The authors have no competing interests to declare that are relevant to the content of this article.

\bibliographystyle{spmpsci}      % mathematics and physical sciences
\bibliography{citations}   % name your BibTeX data base

\end{document}